\documentclass[12pt,thmsa]{article}
\usepackage{amsmath}
\usepackage{graphicx}

\begin{document}

\textbf{The Proof that Maxwell Equations with the 3D E and B }

\textbf{are not Covariant upon the Lorentz Transformations but }

\textbf{upon the Standard Transformations. The New Lorentz }

\textbf{Invariant Field Equations}\bigskip \bigskip

\qquad Tomislav Ivezi\'{c}

\qquad\textit{Ru%
\mbox
{\it{d}\hspace{-.15em}\rule[1.25ex]{.2em}{.04ex}\hspace{-.05em}}er Bo\v
{s}kovi\'{c} Institute, P.O.B. 180, 10002 }

\qquad\textit{Zagreb, Croatia}

\textit{\qquad ivezic@irb.hr\bigskip\medskip}

\noindent In this paper the Lorentz transformations (LT) and the standard
transformations (ST) of the usual Maxwell equations (ME) with the
three-dimensional (3D) vectors of the electric and magnetic fields, $\mathbf{%
E}$ and $\mathbf{B}$ respectively, are examined using both the geometric
algebra and tensor formalisms. Different 4D algebraic objects are used to
represent the usual observer dependent and the new observer independent
electric and magnetic fields. It is found that \emph{the ST of the ME differ
from their LT and consequently that the ME with the 3D }$\mathbf{E}$ \emph{%
and }$\mathbf{B}$ \emph{are not covariant upon the LT but upon the ST}. The
obtained results do not depend on the character of the 4D algebraic objects
used to represent the electric and magnetic fields. The Lorentz invariant
field equations are presented with 1-vectors $E$\ and $B$, bivectors $E_{Hv}$
and $B_{Hv}$\ and the abstract tensors, the 4-vectors $E^{a}$\textbf{\ }and%
\textbf{\ }$B^{a}$. All these quantities are defined without reference
frames, i.e., as absolute quantities. When some basis has been introduced,
they are represented as coordinate-based geometric quantities comprising
both components and a basis. It is explicitly shown that this geometric
approach agrees with experiments, e.g., the Faraday disk, in all relatively
moving inertial frames of reference, which is not the case with the usual
approach with the 3D $\mathbf{E}$ and $\mathbf{B}$ and their ST$\mathbf{.}$
\bigskip

\noindent KEY WORDS: standard and Lorentz transformations of Maxwell
equations.\medskip \bigskip

\noindent \textbf{1. INTRODUCTION\bigskip }

\noindent Recently it is shown in the tensor formalism$^{(1)}$ and the
geometric (Clifford) algebra formalism,$^{(2)}$ that the standard
transformations (ST)$^{(3,4)}$ (see also the well-known textbooks, e.g.
Refs. 5,6) of the three-dimensional (3D) vectors of the electric and
magnetic fields, $\mathbf{E}$ and $\mathbf{B}$ respectively, drastically
differ from the correct Lorentz transformations (LT) of the corresponding 4D
algebraic objects representing the electric and magnetic fields. The
fundamental difference is that in the ST, e.g., the components of the
transformed 3D $\mathbf{E}_{st}^{\prime }$ are expressed by the mixture of
components of the 3D $\mathbf{E}$ and $\mathbf{B}$, and similarly for $%
\mathbf{B}_{st}^{\prime }$. However, the correct LT always transform the 4D
algebraic object representing the electric field only to the electric field,
and similarly for the magnetic field. The results from Refs. 1, 2 are used
here to investigate the LT and the ST of the usual Maxwell equations (ME)
with the 3D $\mathbf{E}$ and $\mathbf{B}$. Different 4D algebraic objects
are used to represent the standard observer dependent and the new observer
independent electric and magnetic fields. First the electric and magnetic
fields are represented by the \emph{observer dependent }1-vectors $E_{f}$
and $B_{f}$ defined in the $\gamma _{0}$ - frame. The usual ME in the
component form are derived in Sec. 2.1. The LT of the ME are considered in
Sec. 2.3. It is explicitly shown in Sec. 2.3., using the correct LT of $%
E_{f} $\ and $B_{f}$, that \emph{the Lorentz transformed ME} \emph{are not
of the same form as the original ones.} This proves that, contrary to the
general opinion, \emph{the usual ME are not Lorentz covariant equations. }In
Sec. 2.4. the ST of the usual ME are considered taking into account the ST
of the components of the 3D $\mathbf{E}$ and $\mathbf{B}$. It is proved that
both the ST of the 3D $\mathbf{E}$ and $\mathbf{B}$ and the ST of the usual
ME have nothing in common with the correct LT. The new Lorentz\ invariant\
field\ equations\ are constructed in Sec. 2.6. in which the electric and
magnetic fields are represented by the 1-vectors $E$\ and $B$ that are \emph{%
defined without reference frames.} The whole consideration is briefly
repeated in the same sections using the \emph{observer dependent} bivectors $%
\mathbf{E}_{H}$ and $\mathbf{B}_{H}$ defined in the $\gamma _{0}$ - frame
and the \emph{coordinate-free} bivectors $E_{Hv}$ and $B_{Hv}$.\ In the
geometric algebra formalism the active LT are used. This geometric approach
is compared with the usual manner in which the ME with the 3D $\mathbf{E}$
and $\mathbf{B}$ are obtained from the covariant ME with $F^{\mu \nu }$,
Sec. 2.2., and from the Lorentz transformed $F^{\prime \mu \nu }$, Sec.
2.5.. In Sec. 3. the whole consideration is performed in the tensor
formalism using the \emph{coordinate-free} 4-vectors $E^{a}$\textbf{\ }and%
\textbf{\ }$B^{a}$ and the \emph{observer dependent} 4-vectors $E_{f}^{a}$
and $B_{f}^{a}$ defined in the $\gamma _{0}$ - frame. In the tensor
formalism the passive LT are used. All quantities in the Lorentz invariant
field equations, with 1-vectors $E$\ and $B$, bivectors $E_{Hv}$ and $B_{Hv}$%
\ and the abstract 4-vectors $E^{a}$\textbf{\ }and\textbf{\ }$B^{a}$ are
geometric, coordinate-free quantities, i.e., the absolute quantities (AQs).
They are defined without reference frames, or, when some basis has been
introduced, they are represented as coordinate-based geometric quantities
(CBGQs) comprising both components and a basis. All such equations are
completely equivalent to the field equations with $F$ (given, e.g. in
Refs.7-9 and discussed in detail in Ref. 10) or with $F^{ab}$ (already
presented, e.g., in Ref. 11). It can be concluded from the consideration
presented in all mentioned sections that the difference between the ST and
the LT of the ME does not depend on the character of the 4D algebraic
objects used to represent the electric and magnetic fields. The comparison
with experiments is given in Sec. 4. and it shows that this geometric
approach agrees with experiments, e.g., the Faraday disk, in all relatively
moving inertial frames of reference, which is not the case with the usual
approach with the 3D $\mathbf{E}$ and $\mathbf{B}$ and their ST$\mathbf{.}$
(The comparison of the geometric approach to special relativity (SR) and of
the standard formulation of SR with experiments that test SR is also given
in detail in Ref. 12.) The summary and conclusions are presented in Sec. 5.
(We note that the great part of the consideration exposed in this paper is
also presented in Ref. 13.)\textbf{\bigskip \medskip }

\noindent \textbf{2. THE PROOF OF THE DIFFERENCE\ BETWEEN\ THE\ LT }

\textbf{AND THE\ ST\ OF\ THE ME\ USING THE GEOMETRIC\ }

\textbf{ALGEBRA APPROACH}\bigskip

For the usual formulation of electrodynamics with the Clifford multivectors,
see, e.g., Refs. 7-9. In Refs. 7-9 the electromagnetic field is represented
by a bivector-valued function $F=F(x)$ on the spacetime. The source of the
field is the electromagnetic current $j$ which is a 1-vector field and the
gradient operator $\partial $ is also 1-vector. A single field equation for $%
F$ is first given by M. Riesz$^{(14)}$ as
\begin{equation}
\partial F=j/\varepsilon _{0}c,\quad \partial \cdot F+\partial \wedge
F=j/\varepsilon _{0}c.  \label{MEF}
\end{equation}
The trivector part is identically zero in the absence of magnetic charge.
The geometric (Clifford) product is written by simply juxtaposing
multivectors $AB$. The dot $``\cdot "$ and wedge $``\wedge "$ in (\ref{MEF})
denote the inner and outer products respectively. All quantities in (\ref
{MEF}) are AQs. Thence they are independent of the reference frame and the
chosen system of coordinates in that frame. Consequently the equation (\ref
{MEF}) is a Lorentz invariant field equation. In the geometric algebra
formalism (as in the tensor formalism as well) one mainly deals either with
4D AQs, e.g., the Clifford multivector $F$ (the abstract tensor $F^{ab}$)
or, when some basis has been introduced, with CBGQs that comprise both
components and a basis. The SR that exclusively deals with AQs or,
equivalently, with CBGQs, can be called the invariant SR.$^{(11,12,10,15)}$
The reason for this name is that upon the passive LT any 4D CBGQ remains
unchanged. The invariance of some 4D CBGQ upon the passive LT reflects the
fact that such mathematical, invariant, geometric 4D quantity represents
\emph{the same physical object} for relatively moving observers. \emph{It is
taken in the invariant SR that such 4D geometric quantities are well-defined
not only mathematically but also experimentally, as measurable quantities
with real physical meaning. Thus they have an independent physical reality. }

In the usual geometric algebra formalism, e.g., Refs. 7, 8, 9, instead of to
work only with such \emph{observer independent quantities} one introduces
(in order to get a more familiar form for (\ref{MEF})) a space-time split
and the relative vectors in the $\gamma _{0}$ - frame, i.e., a particular
time-like direction $\gamma _{0}$ is singled out. $\gamma _{0}$ is tangent
to the world line of an observer at rest in the $\gamma _{0}$ - frame.

(The generators of the spacetime algebra are four basis vectors $\gamma
_{\mu },\mu =0...3,$ satisfying $\gamma _{\mu }\cdot \gamma _{\nu }=\eta
_{\mu \nu }=diag(+---).$ They form the standard basis $\left\{ \gamma _{\mu
}\right\} $. This basis is a right-handed orthonormal frame of vectors in
the Minkowski spacetime $M^{4}$ with $\gamma _{0}$ in the forward light
cone. The $\gamma _{k}$ ($k=1,2,3$) are spacelike vectors. The $\gamma _{\mu
}$ generate by multiplication a complete basis for spacetime algebra: $%
1,\gamma _{\mu },\gamma _{\mu }\wedge \gamma _{\nu },\gamma _{\mu }\gamma
_{5,}\gamma _{5}$ (16 independent elements). $\gamma _{5}$ is the
pseudoscalar for the frame $\left\{ \gamma _{\mu }\right\} .$ It is worth
noting that the standard basis corresponds, in fact, to the specific system
of coordinates, i.e., to Einstein's system of coordinates. In the Einstein
system of coordinates the Einstein synchronization$^{(4)}$ of distant clocks
and Cartesian space coordinates $x^{i}$ are used in the chosen inertial
frame of reference. However \emph{different systems of coordinates of an
inertial frame of reference are allowed and they are all equivalent in the
description of physical phenomena. }For example, in Ref. 11 two very
different, but completely equivalent systems of coordinates, the Einstein
system of coordinates and ``radio'' (``r'') system of coordinates, are
exposed and exploited throughout the paper. In this paper, for the sake of
brevity and of clearness of the whole exposition, we shall work only with
the standard basis $\left\{ \gamma _{\mu }\right\} $, but remembering that
the approach with 4D quantities that are defined without reference frames
holds for any choice of the basis.)

The bivector field $F$ is decomposed in the $\gamma _{0}$ - frame into
electric and magnetic parts using different algebraic objects to represent
these fields. The explicit appearance of $\gamma _{0}$ in these expressions
implies that \emph{the space-time split is observer dependent} and thus all
quantities obtained by the space-time split in the $\gamma _{0}$ - frame are
\emph{observer dependent quantities}. In Refs. 7,8 the \emph{observer
independent} $F$ field from (\ref{MEF}) is expressed in terms of \emph{%
observer dependent quantities,} i.e., as the sum of a relative vector $%
\mathbf{E}_{H}$ and a relative bivector $\gamma _{5}\mathbf{B}_{H}$
\begin{align}
F& =\mathbf{E}_{H}+c\gamma _{5}\mathbf{B}_{H}\mathbf{,\quad E}_{H}=(F\cdot
\gamma _{0})\gamma _{0}=(1/2)(F-\gamma _{0}F\gamma _{0}),  \notag \\
\gamma _{5}\mathbf{B}_{H}& =(1/c)(F\wedge \gamma _{0})\gamma
_{0}=(1/2c)(F+\gamma _{0}F\gamma _{0}).  \label{FB}
\end{align}
(The subscript $H$ is for ``Hestenes.'') Both $\mathbf{E}_{H}$ and $\mathbf{B%
}_{H}$ are, in fact, bivectors. Similarly in Ref. 9 $F$ is decomposed in
terms of \emph{observer dependent quantities}, 1-vector $\mathbf{E}_{J}$ and
a bivector $\mathbf{B}_{J}$ (the subscript $J$ is for ``Jancewicz'') as $%
F=\gamma _{0}\wedge \mathbf{E}_{J}-c\mathbf{B}_{J},$ where $\mathbf{E}%
_{J}=F\cdot \gamma _{0}$ and $\mathbf{B}_{J}=-(1/c)(F\wedge \gamma
_{0})\gamma _{0}.$ The $F$ field can be also decomposed in terms of other
algebraic objects; the \emph{observer dependent }electric and magnetic parts
of $F$ are represented with 1-vectors that are denoted as $E_{f}$ and $B_{f}$
(see also Refs. 2, 15). The physical description with 1-vectors $E_{f}$ and $%
B_{f}$ is simpler but \emph{completely equivalent }to the description with
the bivectors $\mathbf{E}_{H},$ $\mathbf{B}_{H}$, Refs. 7,8, or with
1-vector $\mathbf{E}_{J}$ and a bivector $\mathbf{B}_{J}$, Ref. 9. Such
decomposition of $F$ is not only simpler but also much closer to the
classical representation of the electric and magnetic fields by the 3D
vectors $\mathbf{E}$ and $\mathbf{B}$ than those used in Refs. 7, 8, 9. Thus
\begin{align}
F& =E_{f}\wedge \gamma _{0}+c(\gamma _{5}B_{f})\cdot \gamma _{0},  \notag \\
E_{f}& =F\cdot \gamma _{0},\ B_{f}=-(1/c)\gamma _{5}(F\wedge \gamma _{0}).
\label{ebg}
\end{align}

Having at our disposal different decompositions of $F$ into \emph{observer
dependent quantities} we proceed to present the difference between the ST
and the LT of the ME using the decomposition (\ref{ebg}) and only briefly
the decomposition (\ref{FB}). We shall not deal with the decomposition of $F$
into $\mathbf{E}_{J}$ and $\mathbf{B}_{J}$ from Ref. 9 since both the
procedure and the results are completely the same as with the decompositions
(\ref{ebg}) and (\ref{FB}).\bigskip \medskip

\noindent \textbf{2.1. The Field Equations in the} $\gamma _{0}$ - \textbf{%
Frame.} \textbf{The Maxwell Equations}\bigskip

When (\ref{ebg}) is introduced into the field equation for $F$, Eq. (\ref
{MEF}), we find
\begin{align}
\partial \lbrack (F\cdot \gamma _{0})\wedge \gamma _{0}+(F\wedge \gamma
_{0})\cdot \gamma _{0}]& =j/\varepsilon _{0}c  \notag \\
\partial (E_{f}\wedge \gamma _{0}+c(\gamma _{5}B_{f})\cdot \gamma _{0})&
=j/\varepsilon _{0}c.  \label{eqfi}
\end{align}
The equations (\ref{eqfi}) can be now written as coordinate-based geometric
equations (CBGEs) in the standard basis $\left\{ \gamma _{\mu }\right\} $
and the second equation becomes
\begin{align}
\{\partial _{\alpha }[\delta _{\quad \mu \nu }^{\alpha \beta }E_{f}^{\mu
}(\gamma _{0})^{\nu }+c\varepsilon ^{\alpha \beta \mu \nu }(\gamma
_{0})_{\mu }B_{f,\nu }]-(j^{\beta }/c\varepsilon _{0})\}\gamma _{\beta }+&
\notag \\
\partial _{\alpha }[\delta _{\quad \mu \nu }^{\alpha \beta }(\gamma
_{0})^{\mu }cB_{f}^{\nu }+\varepsilon ^{\alpha \beta \mu \nu }(\gamma
_{0})_{\mu }E_{f,\nu }]\gamma _{5}\gamma _{\beta }& =0,  \label{cl}
\end{align}
where $\gamma _{0}=(\gamma _{0})^{\mu }\gamma _{\mu }$ with $(\gamma
_{0})^{\mu }=(1,0,0,0)$ and
\begin{align}
E_{f}& =E_{f}^{\mu }\gamma _{\mu }=0\gamma _{0}+F^{i0}\gamma _{i},  \notag \\
B_{f}& =B_{f}^{\mu }\gamma _{\mu }=0\gamma _{0}+(-1/2c)\varepsilon
^{0kli}F_{kl}\gamma _{i}.  \label{gnl}
\end{align}
Thence the components of $E_{f}$ and $B_{f}$ in the $\left\{ \gamma _{\mu
}\right\} $ basis are
\begin{equation}
E_{f}^{i}=F^{i0},\quad B_{f}^{i}=(-1/2c)\varepsilon ^{0kli}F_{kl}.
\label{sko}
\end{equation}
The relation (\ref{sko}) is nothing else than the standard identification of
the components $F^{\mu \nu }$ with the components of the 3D vectors $\mathbf{%
E}$ and $\mathbf{B,}$ see, e.g., Refs. 1,2. (It is worth noting that
Einstein's fundamental work$^{(16)}$ is the earliest reference on covariant
electrodynamics and on the identification of some components of $F^{\alpha
\beta }$ with the components of the 3D $\mathbf{E}$ and $\mathbf{B.}$) We
see that in the $\gamma _{0}$ - frame $E_{f}$ \emph{and} $B_{f}$ \emph{do
not have the temporal components} $E_{f}^{0}=B_{f}^{0}=0$. Thus $E_{f}$ and $%
B_{f}$ actually refer to the 3D subspace orthogonal to the specific timelike
direction $\gamma _{0}.$ Notice that we can select a particular, but
otherwise arbitrary, inertial frame of reference as the $\gamma _{0}$ -
frame, to which we shall refer as the frame of our ``fiducial'' observers
(for this name see Ref. 17). The subscript $``f"$ in the above relations
stands for ``fiducial'' and denotes the explicit dependence of these
quantities on the $\gamma _{0}$ -, i.e., ``fiducial'' - observer.

Using that $E_{f}^{0}=B_{f}^{0}=0$ and $(\gamma _{0})^{\mu }=(1,0,0,0)$ the
equation (\ref{cl}) becomes
\begin{align}
(\partial _{k}E_{f}^{k}-j^{0}/c\varepsilon _{0})\gamma _{0}+(-\partial
_{0}E_{f}^{i}+c\varepsilon ^{ijk0}\partial _{j}B_{fk}-j^{i}/c\varepsilon
_{0})\gamma _{i}+&  \notag \\
(-c\partial _{k}B_{f}^{k})\gamma _{5}\gamma _{0}+(c\partial
_{0}B_{f}^{i}+\varepsilon ^{ijk0}\partial _{j}E_{fk})\gamma _{5}\gamma _{i}&
=0.  \label{MEC}
\end{align}
The first part (with $\gamma _{\alpha }$) in Eq. (\ref{MEC}) is from the
1-vector part of Eq. (\ref{eqfi}), i.e., Eq. (\ref{cl}), whereas the second
one (with $\gamma _{5}\gamma _{\alpha }$) is from the trivector
(pseudovector) part of Eq. (\ref{eqfi}), i.e., Eq. (\ref{cl}). Both parts in
Eq. (\ref{MEC}) are written as CBGEs in the standard basis $\left\{ \gamma
_{\mu }\right\} $ and cannot be further simplified as geometric equations.
In the first part (with $\gamma _{\alpha }$) in Eq. (\ref{MEC}) one
recognizes \emph{two} \emph{Maxwell equations} in the \emph{component form},
the Gauss law for the electric field (the first bracket, with $\gamma _{0}$)
and the Amp\`{e}re-Maxwell law (the second bracket, with $\gamma _{i}$).
Similarly from the second part (with $\gamma _{5}\gamma _{\alpha }$) in Eq. (%
\ref{MEC}) we recognize the \emph{component form} of another \emph{two
Maxwell equations}, the Gauss law for the magnetic field (with $\gamma
_{5}\gamma _{0}$) and Faraday's law (with $\gamma _{5}\gamma _{i}$).

The whole procedure can be repeated using the decomposition of $F$, Eq. (\ref
{FB}), into the bivectors $\mathbf{E}_{H},$ $\mathbf{B}_{H}$ as in Refs.
7,8. We shall quote only the results (the complete derivation is given in
Ref. 13). When the decomposition (\ref{FB}) is substituted into Eq. (\ref
{MEF}) we find
\begin{equation}
\partial (\mathbf{E}_{H}+c\gamma _{5}\mathbf{B}_{H})=j/\varepsilon _{0}c.
\label{H1}
\end{equation}
All quantities in Eq. (\ref{H1}) can be written as CBGQs in the standard
basis $\left\{ \gamma _{\mu }\right\} $ (see also Refs. 2, 13). Thus $%
\mathbf{E}_{H}=F^{i0}\gamma _{i}\wedge \gamma _{0},$ $\mathbf{B}%
_{H}=(1/2c)\varepsilon ^{kli0}F_{kl}\gamma _{i}\wedge \gamma _{0}.$ Both
bivectors $\mathbf{E}_{H}$ and $\mathbf{B}_{H}$ are parallel to $\gamma _{0}$%
, that is, it holds that $\mathbf{E}_{H}\wedge \gamma _{0}=\mathbf{B}%
_{H}\wedge \gamma _{0}=0$. When written in terms of components (e.g., $(%
\mathbf{E}_{H})^{\mu \nu }=\gamma ^{\nu }\cdot (\gamma ^{\mu }\cdot \mathbf{E%
}_{H})=(\gamma ^{\nu }\wedge \gamma ^{\mu })\cdot \mathbf{E}_{H}$) one finds
that $\mathbf{E}_{H}=(E_{H})^{i0}\gamma _{i}\wedge \gamma _{0}=E^{i}\gamma
_{i}\wedge \gamma _{0},$ $\mathbf{B}_{H}=(B_{H})^{i0}\gamma _{i}\wedge
\gamma _{0}=B^{i}\gamma _{i}\wedge \gamma _{0}$. Thus it holds that $(%
\mathbf{E}_{H})^{ij}=(\mathbf{B}_{H})^{ij}=0$. \emph{Multiplying} Eq. (\ref
{H1}) \emph{by} $\gamma _{0}$ and using the above expressions for $\mathbf{E}%
_{H},$ $\mathbf{B}_{H}$ we write the resulting equation as a CBGE
\begin{align}
(\partial _{k}E^{k}-j^{0}/c\varepsilon _{0})+(\partial
_{0}E^{i}-c\varepsilon ^{ijk0}\partial _{j}B_{k}+j^{i}/c\varepsilon
_{0})(\gamma _{i}\wedge \gamma _{0})+&  \notag \\
(c\partial _{k}B^{k})\gamma _{5}+(c\partial _{0}B^{i}+\varepsilon
^{ijk0}\partial _{j}E_{k})\gamma _{5}(\gamma _{i}\wedge \gamma _{0})& =0.
\label{H2}
\end{align}
The equation (\ref{H2}) is exactly the same as the equations obtained in the
geometric algebra formalism, e.g., the equations (8.5) and (8.6a-8.6d) in
the first of Ref. 7, now written as a CBGE. Eq. (\ref{H2}) encodes all four
ME in the component form in the same way as it happens with the equation (%
\ref{MEC}). It is worth noting that this step, the multiplication of Eq. (%
\ref{H1}) by $\gamma _{0}$, in order to get the usual ME, is unnecessary in
the formulation with 1-vectors $E_{f}$ and $B_{f}.$ This shows that the
approach with 1-vectors $E_{f}$ and $B_{f}$ is simpler than the approach
with bivectors $\mathbf{E}_{H}$ and $\mathbf{B}_{H}$ and also it is much
closer to the classical formulation of electromagnetism with the 3D vectors $%
\mathbf{E}$ and $\mathbf{B}.\bigskip \medskip $

\noindent \textbf{2.2. The Comparison of the usual Covariant Approach and }

\textbf{the Geometric Approach,} \textbf{I}\bigskip

Let us now examine the difference between the usual covariant approach,
e.g., Refs. 5,6, and the above geometric approach. The covariant approach
deals with the \emph{component form} (implicitly taken in the standard basis
$\left\{ \gamma _{\mu }\right\} $) of the ME with $F^{\alpha \beta }$ and
its dual $^{\ast }F^{\alpha \beta }$
\begin{equation}
\partial _{\alpha }F^{a\beta }=j^{\beta }/\varepsilon _{0}c,\quad \partial
_{\alpha }\ ^{\ast }F^{\alpha \beta }=0,  \label{maxco}
\end{equation}
where $^{\ast }F^{\alpha \beta }=(1/2)\varepsilon ^{\alpha \beta \gamma
\delta }F_{\gamma \delta }$. (Almost always in the usual covariant
approaches to SR one considers \emph{only the components} of the geometric
quantities taken in the $\left\{ \gamma _{\mu }\right\} $ basis and thus not
the whole tensor. However the components are coordinate quantities and they
do not contain the whole information about the physical quantity.) In order
to get the component form of the ME with the 3D\textbf{\ }$\mathbf{E}$%
\textbf{\ }and $\mathbf{B}$
\begin{eqnarray}
\partial _{k}E_{k}-j^{0}/c\varepsilon _{0} &=&0,\quad -\partial
_{0}E_{i}+c\varepsilon _{ikj}\partial _{j}B_{k}-j^{i}/c\varepsilon _{0}=0,
\notag \\
\partial _{k}B_{k} &=&0,\quad c\partial _{0}B_{i}+\varepsilon _{ikj}\partial
_{j}E_{k}=0  \label{j3}
\end{eqnarray}
from Eq. (\ref{maxco}) one simply makes \emph{the identification of six
independent components of} $F^{\mu \nu }$ \emph{with three components} $%
E_{i} $ \emph{and three components} $B_{i}$
\begin{equation}
E_{i}=F^{i0},\quad B_{i}=(1/2c)\varepsilon _{ikl}F_{lk}.  \label{sko1}
\end{equation}
(The components of the 3D fields $\mathbf{E}$ and $\mathbf{B}$ are written
with lowered (generic) subscripts, since they are not the spatial components
of the 4D quantities. This refers to the third-rank antisymmetric $%
\varepsilon $ tensor too. The super- and subscripts are used only on the
components of the 4D quantities.) Then the 3D $\mathbf{E}$ and $\mathbf{B}$,
as \emph{geometric quantities in the 3D space}, are constructed from these
six independent components of $F^{\mu \nu }$ and \emph{the unit 3D vectors }$%
\mathbf{i},$ $\mathbf{j},$ $\mathbf{k,}$ e.g., $\mathbf{E=}F^{10}\mathbf{i}%
+F^{20}\mathbf{j}+F^{30}\mathbf{k}$. The usual ME with the 3D $\mathbf{E}$
and $\mathbf{B}$ are obtained from Eq. (\ref{j3}) and so constructed 3D $%
\mathbf{E}$ and $\mathbf{B}$ as
\begin{eqnarray}
\nabla \mathbf{E}(\mathbf{r},t) &=&\rho (\mathbf{r},t)/\varepsilon
_{0},\quad \nabla \times \mathbf{E}(\mathbf{r},t)=-\partial \mathbf{B}(%
\mathbf{r},t)/\partial t  \notag \\
\nabla \mathbf{B}(\mathbf{r},t) &=&0,\quad \nabla \times \mathbf{B}(\mathbf{r%
},t)=(1/\varepsilon _{0}c^{2})\mathbf{j}(\mathbf{r},t)+(1/c^{2})\partial
\mathbf{E}(\mathbf{r},t)/\partial t.  \label{max}
\end{eqnarray}
Such usual procedure has a number of disadvantages. They are:

\noindent i) The covariant ME (\ref{maxco}) are written in the component
form and these components are taken in the Einstein system of coordinates,
whereas the field equation (\ref{MEF}) is written with AQs, i.e., it is
independent of the reference frame and of the chosen system of coordinates
in that frame. When Eq. (\ref{MEF}) is written as a CBGE in the $\gamma _{0}$
- frame with the $\left\{ \gamma _{\mu }\right\} $ basis and when \emph{only}
the components are taken then Eq. (\ref{MEF}) becomes Eq. (\ref{maxco}).

\noindent ii) It is considered by the identification (\ref{sko1}) that $%
E_{i} $ and $B_{i}$ are the primary quantities for the whole
electromagnetism and that the components $F^{\alpha \beta }$ are derived
from and determined with $E_{i}$ and $B_{i}$. But the components $F^{\alpha
\beta }$ are determined as the solutions of the field equations (\ref{maxco}%
) for the given sources and, in principle, they are not in any obvious
relation with $E_{i}$ and $B_{i}$, which are the solutions of Eq. (\ref{j3}%
). It is shown in Ref. 10 that the whole electromagnetism can be formulated
exclusively by the well-defined geometric 4D quantity, the Faraday bivector $%
F$, without even mentioning the 3D\textbf{\ }$\mathbf{E}$ and $\mathbf{B}$
or the 4D electromagnetic potentials (which are gauge dependent). Thus $F$
is the primary quantity and not the 3D\textbf{\ }$\mathbf{E}$ and $\mathbf{B}
$, or the potentials.

\noindent iii) The simple identification (\ref{sko1}) of the components $%
E_{i}$ and $B_{i}$ with the components of $F^{\alpha \beta }$ is not a
permissible tensor operation; permissible tensor operations with components
of tensors produce components of new tensors, for example: a) multiplication
by a scalar field \ b) addition of components of two tensors\ c) contraction
on a pair of indices, ... .

\noindent iv) Such identification\ of the components of the 3D $\mathbf{E}$\
and $\mathbf{B}$\textbf{\ }with components of $F^{\mu \nu }$ is dependent on
the chosen system of coordinates. In the usual covariant approaches the
standard basis $\left\{ \gamma _{\mu }\right\} $ is implicitly assumed.%
\textbf{\ }However\textbf{\ }the identification (\ref{sko1}) is meaningless,
e.g., in the ``r'' system of coordinates, the $\left\{ r_{\mu }\right\} $
basis,$^{(11)}$ in which only the Einstein synchronization is replaced by an
asymmetric synchronization, the ''radio'' synchronization.$^{(11)}$ Then $%
F_{r}^{10}=F^{10}+F^{12}+F^{13}$, which means that by the relation (\ref
{sko1}) $E_{1r}=F_{r}^{10}$ the component $E_{1r}$ in the $\left\{ r_{\mu
}\right\} $ basis is expressed as the combination of $E_{i}$ and $B_{i}$
components from the $\left\{ \gamma _{\mu }\right\} $ basis, $%
E_{1r}=E_{1}-B_{3}+B_{2}$, see Ref. 11.

\noindent v) $E_{i}$ and $B_{i}$ in Eq. (\ref{j3}) are the components of
vectors defined on the 3D space while $F^{\alpha \beta }$ are the components
of tensor defined on the 4D spacetime. Thence when forming the geometric
quantities the components of the 4D quantity would need to be multiplied
with the unit vectors $\gamma _{i}$ from the 4D spacetime and not with the
unit vectors $\mathbf{i},$ $\mathbf{j},$ $\mathbf{k}$ from the 3D space.

On the other hand in the above geometric approach the mapping between $F$
and 1-vectors $E_{f}$, $B_{f}$, or bivectors $\mathbf{E}_{H}$, $\mathbf{B}%
_{H}$, given by the equations (\ref{ebg}) and (\ref{FB}) respectively, is
performed by a correct mathematical procedure and all quantities are defined
on the same 4D spacetime. Instead of Eq. (\ref{j3}) that contains a
combination of quantities (components) from the 4D spacetime ($\partial
_{\mu }$, $j^{\mu }$) and from the 3D space ($E_{i}$, $B_{i}$, $\varepsilon
_{ikj}$), we have the CBGEs (\ref{MEC}) and (\ref{H2}) in the geometric
approach, which contain \emph{only} components $E_{f}^{\mu }$, $B_{f}^{\mu }$
and $(\mathbf{E}_{H})^{\mu \nu }$, $(\mathbf{B}_{H})^{\mu \nu }$ of the
well-defined 4D quantities $E_{f}$, $B_{f}$, and $\mathbf{E}_{H}$, $\mathbf{B%
}_{H}$. Similarly instead of the usual ME (\ref{max}) with \emph{geometric
quantities from the 3D space} $\mathbf{E}$ \emph{and} $\mathbf{B}$ we have
the ME (\ref{eqfi}) and (\ref{H1}) with \emph{geometric quantities from the
4D spacetime} $E_{f}$, $B_{f}$, \emph{and} $\mathbf{E}_{H}$, $\mathbf{B}_{H}$%
. However it has to be noted that the decompositions (\ref{ebg}) and (\ref
{FB}) still have some disadvantages. In Eqs. (\ref{ebg}) and (\ref{FB}) the
\emph{observer independent} 4D quantity $F$ is decomposed into the \emph{%
observer dependent} 4D quantities $E_{f}$, $B_{f}$, or $\mathbf{E}_{H}$, $%
\mathbf{B}_{H}$ by using the space-time split in the $\gamma _{0}$ - frame.
The space-time split in another $\gamma _{0}^{\prime }$ - frame is not
obtained by the LT from that one in the $\gamma _{0}$ - frame. This problem
will be discussed in the subsequent sections and in Secs. 2.6. and 3. we
shall present the new decompositions of $F$ without using the space-time
split. \medskip \bigskip

\noindent \textbf{2.3. The LT of the Maxwell Equations} \bigskip

Let us now apply the active LT upon Eq. (\ref{MEC}), or Eq. (\ref{cl}). We
write Eq. (\ref{MEC}), or Eq. (\ref{cl}), in the form
\begin{equation}
a^{\alpha }\gamma _{\alpha }+b^{\alpha }(\gamma _{5}\gamma _{\alpha })=0.
\label{ab}
\end{equation}
The coefficients $a^{\alpha }$ and $b^{\alpha }$ are clear from Eq. (\ref
{MEC}), or Eq. (\ref{cl}); they are the usual ME in the component form. In
the Clifford algebra formalism, e.g., Refs. 7-9, the LT are considered as
active transformations; the components of, e.g., some 1-vector relative to a
given inertial frame of reference (with the standard basis $\left\{ \gamma
_{\mu }\right\} $) are transformed into the components of a new 1-vector
relative to the same frame (the basis $\left\{ \gamma _{\mu }\right\} $ is
not changed). Furthermore the LT are described with rotors $R,$ $R\widetilde{%
R}=1,$ in the usual way as $p\rightarrow p^{\prime }=Rp\widetilde{R}%
=p^{\prime \mu }\gamma _{\mu }.$ To an observer in the $\left\{ \gamma _{\mu
}\right\} $ frame the vector $p^{\prime }$ appears the same as the vector $p$
appears to an observer in the $\left\{ \gamma _{\mu }^{\prime }\right\} $
frame. For boosts in the direction $\gamma _{1}$ the rotor $R$ is given by
the relation
\begin{equation}
R=(1+\gamma +\gamma \beta \gamma _{0}\gamma _{1})/(2(1+\gamma ))^{1/2},
\label{err}
\end{equation}
$\beta $ is the scalar velocity in units of $c$, $\gamma =(1-\beta
^{2})^{-1/2}.$ Then the LT of Eq. (\ref{eqfi}) are given as
\begin{align}
R\{\partial \lbrack (F\cdot \gamma _{0})\wedge \gamma _{0}+(F\wedge \gamma
_{0})\cdot \gamma _{0}]-j/\varepsilon _{0}c\}\widetilde{R}& =0,  \notag \\
R\{\partial \lbrack E_{f}\wedge \gamma _{0}+c(\gamma _{5}B_{f})\cdot \gamma
_{0}]-j/\varepsilon _{0}c\}\widetilde{R}& =0,  \label{rem}
\end{align}
where $R$ is given by Eq. (\ref{err}). (A coordinate-free form of the LT is
also given in the Clifford algebra formalism in Ref. 15 and in the tensor
formalism in Ref. 11, see also Ref. 18. The form presented in Ref. 15 does
not need to use rotors but, of course, it can be expressed by rotors as
well.) Then the LT of the usual ME (\ref{ab}) are
\begin{equation}
R\{a^{\alpha }\gamma _{\alpha }+b^{\alpha }(\gamma _{5}\gamma _{\alpha })\}%
\widetilde{R}=0.  \label{lab}
\end{equation}
Performing the LT we find the explicit expression for Eq. (\ref{lab}) as
\begin{align}
\gamma _{0}(\gamma a^{0}-\beta \gamma a^{1})+\gamma _{1}(\gamma a^{1}-\beta
\gamma a^{0})+\gamma _{2}a^{2}+\gamma _{3}a^{3}+&  \notag \\
\gamma _{5}\gamma _{0}(\gamma b^{0}-\beta \gamma b^{1})+\gamma _{5}\gamma
_{1}(\gamma b^{1}-\beta \gamma b^{0})+\gamma _{5}\gamma _{2}b^{2}+\gamma
_{5}\gamma _{3}b^{3}& =0.  \label{L}
\end{align}
It can be simply written as
\begin{equation}
a^{\prime \alpha }\gamma _{\alpha }+b^{\prime \alpha }(\gamma _{5}\gamma
_{\alpha })=0,  \label{L1}
\end{equation}
where, e.g., $a^{\prime 0}=\gamma a^{0}-\beta \gamma a^{1}$ and, as it is
said, $a^{\alpha }$ and $b^{\alpha }$ are the usual ME in the component form
given in Eq. (\ref{MEC}), or Eq. (\ref{cl}). This result, Eq. (\ref{L}),
i.e., Eq. (\ref{L1}), is exactly the usual result for the active LT of a
1-vector and of a pseudovector. It is important to note that, e.g., the
Gauss law for the electric field $a^{0}$ does not transform by the LT again
to the Gauss law but to $a^{\prime 0}$, which is a combination of the Gauss
law and a part of the Amp\`{e}re-Maxwell law ($a^{1}$).

The second equation in (\ref{rem}) can be expressed in terms of Lorentz
transformed derivatives and Lorentz transformed 1-vectors $E_{f}$ and $B_{f}$
as
\begin{equation}
\partial ^{\prime }[E_{f}^{\prime }\wedge (v^{\prime }/c)+c(\gamma
_{5}B_{f}^{\prime })\cdot (v^{\prime }/c)]-j^{\prime }/\varepsilon _{0}c=0,
\label{mcr}
\end{equation}
where $\partial ^{\prime }=R\partial \widetilde{R},$ $v^{\prime }/c=R\gamma
_{0}\widetilde{R}=\gamma \gamma _{0}-\beta \gamma \gamma _{1}$ and (see also
Ref. 2) the Lorentz transformed $E_{f}^{\prime }$ is
\begin{align}
E_{f}^{\prime }& =R(F\cdot \gamma _{0})\widetilde{R}=RE_{f}\widetilde{R}%
=R(F^{i0}\gamma _{i})\widetilde{R}=E_{f}^{\prime \mu }\gamma _{\mu }=  \notag
\\
& =-\beta \gamma E_{f}^{1}\gamma _{0}+\gamma E_{f}^{1}\gamma
_{1}+E_{f}^{2}\gamma _{2}+E_{f}^{3}\gamma _{3},  \label{nle}
\end{align}
what is the usual form for the active LT of the 1-vector $E_{f}=E_{f}^{\mu
}\gamma _{\mu }$. Similarly we find that $B_{f}^{\prime }$ is
\begin{align}
B_{f}^{\prime }& =R\left[ -(1/c)\gamma _{5}(F\wedge \gamma _{0})\right]
\widetilde{R}=RB_{f}\widetilde{R}=R\left[ (-1/2c)\varepsilon
^{0kli}F_{kl}\gamma _{i}\right] \widetilde{R}=  \notag \\
& =B_{f}^{\prime \mu }\gamma _{\mu }=-\beta \gamma B_{f}^{1}\gamma
_{0}+\gamma B_{f}^{1}\gamma _{1}+B_{f}^{2}\gamma _{2}+B_{f}^{3}\gamma _{3}.
\label{nlb}
\end{align}
It is worth noting that $E_{f}^{\prime }$ \emph{and} $B_{f}^{\prime }$ \emph{%
are no longer orthogonal to} $\gamma _{0},$ i.e., \emph{they} \emph{have}
\emph{the temporal components} $\neq 0.$ Furthermore \emph{the components} $%
E_{f}^{\mu }$ ($B_{f}^{\mu }$) \emph{transform upon the active LT again to
the components} $E_{f}^{\prime \mu }$ ($B_{f}^{\prime \mu }$) as seen from
Eqs. (\ref{nle}) and (\ref{nlb}); \emph{there is no mixing of components}.
When Eq. (\ref{mcr}) is written in an expanded form as a CBGE in the
standard basis $\left\{ \gamma _{\mu }\right\} $ it takes the form of Eq. (%
\ref{L1}) but now the coefficients $a^{\prime \alpha }$ are written by means
of the Lorentz transformed components $\partial _{k}^{\prime }$, $%
E_{f}^{\prime k}$ and $B_{f}^{\prime k}$ (for simplicity only the term $%
a^{\prime 0}\gamma _{0}$ is presented)
\begin{equation}
a^{\prime 0}\gamma _{0}=\{[\gamma (\partial _{k}^{\prime }E_{f}^{\prime
k})-j^{\prime 0}/c\varepsilon _{0}]+\beta \gamma \lbrack \partial
_{1}^{\prime }E_{f}^{\prime 0}+c(\partial _{2}^{\prime }B_{f3}^{\prime
}-\partial _{3}^{\prime }B_{f2}^{\prime })]\}\gamma _{0},  \label{anu}
\end{equation}
and \emph{it substantially differs in form from the term} $a^{0}\gamma
_{0}=(\partial _{k}E_{f}^{k}-j^{0}/c\varepsilon _{0})\gamma _{0}$ in Eq. (%
\ref{MEC}). As explained above the coefficient $a^{0}$ is the Gauss law for
the electric field written in the component form. It is clear from Eq. (\ref
{anu}) that the LT do not transform the Gauss law into the ``primed'' Gauss
law but into quite different law Eq. (\ref{anu}); $a^{\prime 0}$ contains
the time component $E_{f}^{\prime 0}$ (while $E_{f}^{0}=0$), and also the
new ``Gauss law'' includes the derivatives of the magnetic field. The same
situation happens with other Lorentz transformed terms, which explicitly
shows that \emph{the Lorentz transformed ME} ((\ref{mcr}) \emph{with} (\ref
{anu})) \emph{are not of the same form as the original ones} Eq. (\ref{MEC}%
). This is a fundamental result which reveals that, contrary to the previous
derivations, e.g., Refs. 4,16,5-9, and contrary to the general opinion,
\emph{the usual ME are not Lorentz covariant equations. }The physical
consequences of this achievement will be very important and they will be
carefully examined.

Again as in Sec 2.1. we give only the results for the case when $\mathbf{E}%
_{H},$ $\mathbf{B}_{H}$ are used (all details are given in Ref. 13.) The
relation (\ref{H2}) can be written in the form $a^{0}+a^{i}(\gamma
_{i}\wedge \gamma _{0})+b^{0}\gamma _{5}+b^{i}\gamma _{5}(\gamma _{i}\wedge
\gamma _{0})=0.$ The coefficients $a^{0}$, $a^{i}$ and $b^{0}$, $b^{i}$ are
clear from Eq. (\ref{H2}); they are the usual ME in the component form. As
it is said the usual ME (\ref{H2}) are obtained multiplying Eq. (\ref{H1})
by $\gamma _{0}$. The LT of the resulting equation (after multiplication by $%
\gamma _{0}$) are
\begin{equation}
R\{\gamma _{0}[\partial (\mathbf{E}_{H}+c\gamma _{5}\mathbf{B}%
_{H})-j/\varepsilon _{0}c]\}\widetilde{R}=0.  \label{reg}
\end{equation}
Then after applying the LT upon Eq. (\ref{H2}) we find $a^{0}+R[a^{i}(\gamma
_{i}\wedge \gamma _{0})]\widetilde{R}+b^{0}\gamma _{5}+R[b^{i}\gamma
_{5}(\gamma _{i}\wedge \gamma _{0})]\widetilde{R}=0,$ where, e.g.,

$R[a^{i}(\gamma _{i}\wedge \gamma _{0})]\widetilde{R}=a^{1}(\gamma
_{1}\wedge \gamma _{0})+\gamma \lbrack a^{2}(\gamma _{2}\wedge \gamma
_{0})+a^{3}(\gamma _{3}\wedge \gamma _{0})]$ $-\beta \gamma \lbrack
a^{2}(\gamma _{2}\wedge \gamma _{1})+a^{3}(\gamma _{3}\wedge \gamma _{1})]$,
see Ref. 13. This result is the usual result for the active LT of a
multivector from Eq. (\ref{H2}). The equation (\ref{reg}) can be expressed
in terms of Lorentz transformed derivatives and Lorentz transformed $\mathbf{%
E}_{H}^{\prime }$ and $\mathbf{B}_{H}^{\prime }$ as
\begin{equation}
(v^{\prime }/c)[\partial ^{\prime }(\mathbf{E}_{H}^{\prime }+c\gamma _{5}%
\mathbf{B}_{H}^{\prime })-j^{\prime }/\varepsilon _{0}c]=0,  \label{ehbc}
\end{equation}
where $v^{\prime }/c=R\gamma _{0}\widetilde{R}$, $\partial ^{\prime
}=R\partial \widetilde{R}$, and the Lorentz transformed bivectors are $%
\mathbf{E}_{H}^{\prime }$ and $\mathbf{B}_{H}^{\prime }$. This $\mathbf{E}%
_{H}^{\prime }$ is

\begin{align}
\mathbf{E}_{H}^{\prime }& =R[(F\cdot \gamma _{0})\gamma _{0}]\widetilde{R}=R%
\mathbf{E}_{H}\widetilde{R}=E^{1}\gamma _{1}\wedge \gamma _{0}+\gamma
(E^{2}\gamma _{2}\wedge \gamma _{0}+  \notag \\
& E^{3}\gamma _{3}\wedge \gamma _{0})-\beta \gamma (E^{2}\gamma _{2}\wedge
\gamma _{1}+E^{3}\gamma _{3}\wedge \gamma _{1}),  \label{eh}
\end{align}
where $E^{i}=F^{i0}$ and it is similarly obtained for $\mathbf{B}%
_{H}^{\prime }$, see Refs. 2, 13. $\mathbf{E}_{H}^{\prime }$, Eq. (\ref{eh})
(and also $\mathbf{B}_{H}^{\prime }$) are the familiar forms for the active
LT of bivectors, here $\mathbf{E}_{H}$ and $\mathbf{B}_{H}$. It is worth
noting that $\mathbf{E}_{H}^{\prime }$ and $\mathbf{B}_{H}^{\prime }$, in
contrast to $\mathbf{E}_{H}$ and $\mathbf{B}_{H}$, are not parallel to $%
\gamma _{0}$, i.e., it \emph{does not hold} \emph{that} $\mathbf{E}%
_{H}^{\prime }\wedge \gamma _{0}=\mathbf{B}_{H}^{\prime }\wedge \gamma
_{0}=0 $ and thus \emph{there are} $(\mathbf{E}_{H}^{\prime })^{ij}\neq 0$
\emph{and} $(\mathbf{B}_{H}^{\prime })^{ij}\neq 0.$ Further, as it happens
for $E_{f}$ and $B_{f}$, see Eqs. (\ref{nle}) and (\ref{nlb}), \emph{the
components} $(\mathbf{E}_{H})^{\mu \nu }$ ($(\mathbf{B}_{H})^{\mu \nu }$)
\emph{transform upon the active LT again to the components} $(\mathbf{E}%
_{H}^{\prime })^{\mu \nu }$ ($(\mathbf{B}_{H}^{\prime })^{\mu \nu }$); \emph{%
there is no mixing of components}. \emph{Thus} \emph{by the active LT} $%
\mathbf{E}_{H}$ \emph{transforms to} $\mathbf{E}_{H}^{\prime }$ \emph{and} $%
\mathbf{B}_{H}$ \emph{to }$\mathbf{B}_{H}^{\prime }.$ Actually, as we said,
this is the way in which every bivector transforms upon the active LT. Then
Eq. (\ref{ehbc}) can be written as a CBGE in the standard basis $\left\{
\gamma _{\mu }\right\} $, but for simplicity we only quote the scalar term $%
a^{\prime 0}$
\begin{align}
a^{\prime 0}& =-\beta \gamma \partial _{0}^{\prime }(\mathbf{E}_{H}^{\prime
})^{10}+\gamma \lbrack \partial _{k}^{\prime }(\mathbf{E}_{H}^{\prime
})^{k0}]+\beta \gamma \lbrack \partial _{2}^{\prime }(\mathbf{E}_{H}^{\prime
})^{21}  \notag \\
& +\partial _{3}^{\prime }(\mathbf{E}_{H}^{\prime })^{31}]-(\gamma j^{\prime
0}-\beta \gamma j^{\prime 1})/\varepsilon _{0}c  \label{act}
\end{align}
Comparing $a^{\prime 0}$, Eq. (\ref{act}), with $a^{0}$ from the usual ME (%
\ref{H2}) $a^{0}=\partial _{k}(\mathbf{E}_{H})^{k0}-j^{0}/c\varepsilon _{0}$%
, we again see, as with $E_{f}$ and $B_{f}$, that $a^{\prime 0}$ \emph{%
substantially differs in form from the term} $a^{0}$ in Eq. (\ref{H2}). The
same situation happens with other transformed terms, which shows that \emph{%
the Lorentz transformed ME}, (\ref{ehbc}) \emph{with} (\ref{act}), \emph{are
not of the same form as the original ones,} Eq. (\ref{H2}). This is a
fundamental result which once again reveals that, contrary to the previous
derivations, e.g., Refs. 4, 16, 5-9, and contrary to the generally accepted
belief, \emph{the usual ME are not Lorentz covariant equations.}\bigskip
\medskip

\noindent \textbf{2.4. The ST of the Maxwell equations} \bigskip

In contrast to the correct active LT of $E_{f}$, Eq. (\ref{nle}), and $B_{f}$%
, Eq. (\ref{nlb}), it is wrongly assumed in the usual derivations of the
\emph{the ST for} $E_{st}^{\prime }$ \emph{and} $B_{st}^{\prime }$ (the
subscript $st$ is for standard)\emph{\ that the quantities obtained by the
active LT of} $E_{f}$ \emph{and} $B_{f}$ \emph{are again in the 3D subspace
of the} $\gamma _{0}$ \emph{-} \emph{observer, }see also Ref. 2. This means
that it is wrongly assumed in all usual derivations, e.g., in the Clifford
algebra formalism$^{(7,8,9)}$ (and in the tensor formalism$^{(16,5,6)}$ as
well), that one can again perform the same identification of the transformed
components $F^{\prime \mu \nu }$ with the components of the 3D $\mathbf{E}%
^{\prime }$ and $\mathbf{B}^{\prime }$ as in Eq. (\ref{sko}). Thus it is
taken in Refs. 7, 8, 9 that for the transformed $E_{st}^{\prime }$ and $%
B_{st}^{\prime }$ again hold $E_{st}^{\prime 0}=B_{st}^{\prime 0}=0$ as for $%
E_{f}$ and $B_{f}$
\begin{align}
E_{st}^{\prime }& =(RF\widetilde{R})\cdot \gamma _{0}=F^{\prime }\cdot
\gamma _{0}=F^{\prime i0}\gamma _{i}=E_{st}^{\prime i}\gamma _{i}=  \notag \\
& =E_{f}^{1}\gamma _{1}+(\gamma E_{f}^{2}-\beta \gamma cB_{f}^{3})\gamma
_{2}+(\gamma E_{f}^{3}+\beta \gamma cB_{f}^{2})\gamma _{3},  \label{ce}
\end{align}
where $F^{\prime }=RF\widetilde{R}$, and similarly for $B_{st}^{\prime }$
\begin{align}
B_{st}^{\prime }& =-(1/c)\gamma _{5}(F^{\prime }\wedge \gamma
_{0})=-(1/2c)\varepsilon ^{0kli}F_{kl}^{\prime }\gamma _{i}=B_{st}^{\prime
i}\gamma _{i}=  \notag \\
& B_{f}^{1}\gamma _{1}+(\gamma B_{f}^{2}+\beta \gamma E_{f}^{3}/c)\gamma
_{2}+(\gamma B_{f}^{3}-\beta \gamma E_{f}^{2}/c)\gamma _{3}.  \label{B}
\end{align}
From \emph{the relativistically incorrect transformations} (\ref{ce}) and (%
\ref{B}) one simply finds the transformations of the spatial components $%
E_{st}^{\prime i}$ and $B_{st}^{\prime i}$
\begin{equation}
E_{st}^{\prime i}=F^{\prime i0},\quad B_{st}^{\prime i}=(-1/2c)\varepsilon
^{0kli}F_{kl}^{\prime }.  \label{sk1}
\end{equation}
As can be seen from Eqs. (\ref{ce}) and (\ref{B}), i.e., from Eq. (\ref{sk1}%
), \emph{the transformations for} $E_{st.}^{\prime i}$ \emph{and} $%
B_{st.}^{\prime i}$ \emph{are exactly the ST of components of the 3D vectors}
$\mathbf{E}$ \emph{and} $\mathbf{B}$ that are quoted in almost every
textbook and paper on relativistic electrodynamics. Notice that, in contrast
to the active LT (\ref{nle}) and (\ref{nlb}), \emph{according to the ST} (%
\ref{ce}), \emph{i.e.}, (\ref{sk1}), \emph{the transformed components} $%
E_{st}^{\prime i}$ \emph{are expressed by the mixture of components} $%
E_{f}^{i}$ \emph{and} $B_{f}^{i},$ \emph{and Eq. }(\ref{B})\emph{\ shows
that the same holds for} $B_{st}^{\prime i}$. In all previous treatments of
SR, e.g., Refs. 7-9 (and Refs. 4,5,6,16) the transformations for $%
E_{st.}^{\prime i}$ and $B_{st.}^{\prime i}$ are considered to be the LT of
the 3D electric and magnetic fields. However the above analysis, and Refs.
1,2 as well, show that the transformations for $E_{st.}^{\prime i}$ and $%
B_{st.}^{\prime i}$, Eq. (\ref{sk1}), are derived from \emph{the
relativistically incorrect transformations} (\ref{ce}) and (\ref{B}), which
are not the LT; the LT are given by the relations (\ref{nle}) and (\ref{nlb}%
).

It is also argued in all previous works, starting in the year 1905 with
Einstein's fundamental paper on SR,$^{(4)}$ that the usual ME with the 3D $%
\mathbf{E}$ and $\mathbf{B}$ are Lorentz covariant equations. The relation (%
\ref{mcr}) together with Eq. (\ref{anu}) shows that it is not true; the
Lorentz transformed ME are not of the same form as the original ones. Here
we explicitly show that \emph{in the usual derivations} \emph{the ME remain
unchanged in form not upon the LT but upon some transformations which,
strictly speaking, have nothing to do with the LT of the equation} (\ref
{eqfi}), \emph{i.e., of the ME} (\ref{MEC}). The difference between the
Lorentz transformed ME, given by Eq. (\ref{rem}) or finally by Eq. (\ref{mcr}%
) with Eq. (\ref{anu}) (or by Eq. (\ref{L})) and the equations (given below)
obtained by applying the ST is the same as the difference between the LT of $%
E_{f}$ ($B_{f}$) given by Eqs. (\ref{nle}) ((\ref{nlb})) and their ST given
by Eqs. (\ref{ce}) ((\ref{B})). Thus the ST of the equation (\ref{eqfi}) are
\begin{align}
(R\partial \widetilde{R})\{[(RF\widetilde{R})\cdot \gamma _{0}]\wedge \gamma
_{0}+[(RF\widetilde{R})\wedge \gamma _{0}]\cdot \gamma _{0}\}-(Rj\widetilde{R%
})/\varepsilon _{0}c& =0,  \notag \\
\partial ^{\prime }\{E_{st}^{\prime }\wedge \gamma _{0}+c(\gamma
_{5}B_{st}^{\prime })\cdot \gamma _{0}\}-j^{\prime }/\varepsilon _{0}c& =0,
\label{rtr}
\end{align}
where $E_{st}^{\prime }$ and $B_{st}^{\prime }$ are determined by Eqs. (\ref
{ce}) and (\ref{B}). Notice that, in contrast to the correct LT (\ref{rem})
or (\ref{mcr}), $\gamma _{0}$ \emph{is not transformed in} Eq. (\ref{rtr}).
When this second equation in (\ref{rtr}) is written as a CBGE in the
standard basis $\left\{ \gamma _{\mu }\right\} $ it becomes
\begin{align}
(\partial _{k}^{\prime }E_{st}^{\prime k}-j^{\prime 0}/c\varepsilon
_{0})\gamma _{0}+(-\partial _{0}^{\prime }E_{st}^{\prime i}+c\varepsilon
^{ijk0}\partial _{j}^{\prime }B_{st,k}^{\prime }-j^{\prime i}/c\varepsilon
_{0})\gamma _{i}+&  \notag \\
(-c\partial _{k}^{\prime }B_{st}^{\prime k})\gamma _{5}\gamma
_{0}+(c\partial _{0}^{\prime }B_{st}^{\prime i}+\varepsilon ^{ijk0}\partial
_{j}^{\prime }E_{st,k}^{\prime })\gamma _{5}\gamma _{i}& =0.  \label{EBC}
\end{align}
The equation (\ref{EBC}) is of the same form as the original ME (\ref{MEC})
but \emph{the electric and magnetic fields are not transformed by the LT
than by the ST}. Therefore, as can be seen from Eq. (\ref{rtr}) (together
with Eqs. (\ref{ce}) and (\ref{B})), \emph{the equation }(\ref{EBC})\emph{\
is not the LT of the original ME} (\ref{MEC}); \emph{the LT of the ME} (\ref
{MEC}) \emph{are the equations} (\ref{mcr}) \emph{with} (\ref{anu}) \emph{%
(i.e.,} Eq. (\ref{L})) where the Lorentz transformed electric and magnetic
fields are given by the relations (\ref{nle}) and (\ref{nlb}).

Let us discuss the ST in the formulation with $\mathbf{E}_{H}$ and $\mathbf{B%
}_{H}$. As can be easily shown, see also Ref. 2, \emph{the ST for} $\mathbf{E%
}_{H,st}^{\prime }$ \emph{and} $\mathbf{B}_{H,st}^{\prime }$ \emph{are
derived wrongly assuming that the quantities obtained by the active LT of} $%
\mathbf{E}_{H}$ \emph{and} $\mathbf{B}_{H}$ \emph{are again parallel to} $%
\gamma _{0}$\emph{, i.e., that again holds} $\mathbf{E}_{H}^{\prime }\wedge
\gamma _{0}=\mathbf{B}_{H}^{\prime }\wedge \gamma _{0}=0$ and consequently
that $(\mathbf{E}_{H,st}^{\prime })^{ij}=(\mathbf{B}_{H,st}^{\prime
})^{ij}=0.$ Thence, in contrast to the correct LT of $\mathbf{E}_{H}$ (Eq. (%
\ref{eh})) (and $\mathbf{B}_{H}$), it is taken in the usual derivations
(Ref. 7, Space-Time Algebra (eq. (18.22)), New Foundations for Classical
Mechanics (Ch. 9 eqs. (3.51a,b)), Ref. 8 (Ch. 7.1.2 eq. (7.33))) that
\begin{align}
\mathbf{E}_{H,st}^{\prime }& =(F^{\prime }\cdot \gamma _{0})\gamma
_{0}=(E_{H,st}^{\prime })^{i0}\gamma _{i}\wedge \gamma _{0}=E_{st}^{\prime
i}\gamma _{i}\wedge \gamma _{0}=  \notag \\
& E^{1}\gamma _{1}\wedge \gamma _{0}+(\gamma E^{2}-\beta \gamma
cB^{3})\gamma _{2}\wedge \gamma _{0}+(\gamma E^{3}+\beta \gamma
cB^{2})\gamma _{3}\wedge \gamma _{0},  \label{es}
\end{align}
where $F^{\prime }=RF\widetilde{R}$, and similarly for $\mathbf{B}%
_{H,st}^{\prime }$, see Ref. 2. The relation (\ref{es}) (and that one for $%
\mathbf{B}_{H,st}^{\prime }$) immediately gives the familiar expressions for
the ST of the 3D vectors $\mathbf{E}$ and $\mathbf{B.}$ Now, in contrast to
the correct LT of $\mathbf{E}_{H}$ (Eq. (\ref{eh})) (and $\mathbf{B}_{H}$),
\emph{the components} \emph{of the transformed }$\mathbf{E}_{H,st}^{\prime }$
\emph{are expressed by the mixture of components} $E^{i}$ \emph{and} $B^{i},$
\emph{and the same holds for} $\mathbf{B}_{H,st}^{\prime }$. The ST of Eq. (%
\ref{H1}) (after multiplication by $\gamma _{0}$) are given as
\begin{equation}
\gamma _{0}[\partial ^{\prime }(\mathbf{E}_{H,st}^{\prime }+c\gamma _{5}%
\mathbf{B}_{H,st}^{\prime })-j^{\prime }/\varepsilon _{0}c]=0,  \label{gc}
\end{equation}
where $\mathbf{E}_{H,st}^{\prime }$ is determined by Eq. (\ref{es}) (and
similarly for $\mathbf{B}_{H,st}^{\prime }$). Notice again that, in contrast
to the correct LT (\ref{reg}) or (\ref{ehbc}), $\gamma _{0}$ \emph{is not
transformed in} Eq. (\ref{gc}), as it is not transformed in the ST $\mathbf{E%
}_{H,st}^{\prime }$, Eq. (\ref{es}) (and $\mathbf{B}_{H,st}^{\prime }$).
When Eq. (\ref{gc}) is written as a CBGE in the standard basis $\left\{
\gamma _{\mu }\right\} $ it becomes $(\partial _{k}^{\prime }E_{st}^{\prime
k}-j^{\prime 0}/c\varepsilon _{0})+(\partial _{0}^{\prime }E_{st}^{\prime
i}-c\varepsilon ^{ijk0}\partial _{j}^{\prime }B_{st,k}^{\prime }+j^{\prime
i}/c\varepsilon _{0})(\gamma _{i}\wedge \gamma _{0})+$ $(c\partial
_{k}^{\prime }B_{st}^{\prime k})\gamma _{5}+(c\partial _{0}^{\prime
}B_{st}^{\prime i}+\varepsilon ^{ijk0}\partial _{j}^{\prime
}E_{st,k}^{\prime })\gamma _{5}(\gamma _{i}\wedge \gamma _{0})=0.$ This
equation is of the same form as the original ME (\ref{H2}) but the bivectors
$\mathbf{E}_{H}$ and $\mathbf{B}_{H}$ representing the electric and magnetic
fields are not transformed by the LT than by the ST. As seen from Eq. (\ref
{gc}) \emph{this equation is not the LT of the original ME} (\ref{H2});
\emph{the LT of the ME} (\ref{H2}) \emph{is the equation} (\ref{ehbc}) \emph{%
with} (\ref{act}).\medskip \bigskip

\noindent \textbf{2.5. The Comparison of the usual Covariant Approach and }

\textbf{the Geometric Approach,} \textbf{II}\bigskip

In the usual covariant approach, e.g., Refs. 5,6, one transforms by the
passive LT the covariant ME (\ref{maxco}) and finds $\partial _{\alpha
}^{\prime }F^{\prime a\beta }=j^{\prime \beta }/\varepsilon _{0}c,$ $%
\partial _{\alpha }^{\prime }\ ^{\ast }F^{\prime \alpha \beta }=0$. (Upon
the passive LT the set of components, e.g., $j^{\mu }$ from the $S$ frame
transform to $j^{\prime \mu }$ in the relatively moving inertial frame of
reference $S^{\prime }$, $j^{\prime \mu }=L_{\ \nu }^{\mu }j^{\nu }$, where
(for the boost in the $\gamma _{1}$ direction) $L_{\ 0}^{0}=L_{\
1}^{1}=\gamma $, $L_{\ 1}^{0}=L_{\ 0}^{1}=-\beta \gamma $, $L_{\ 2}^{2}=L_{\
3}^{3}=1$ and all other components are zero.) Then the same identification
as in Eq. (\ref{sko1}) is assumed to hold for the transformed components $%
E_{i}^{\prime }$ and $B_{i}^{\prime }$
\begin{equation}
E_{i}^{\prime }=F^{\prime i0},\quad B_{i}^{\prime }=(1/2c)\varepsilon
_{ikl}F_{lk}^{\prime },  \label{sko2}
\end{equation}
e.g., $F^{\prime 20}=\gamma F^{20}-\beta \gamma F^{21}$, which yields (by
Eqs. (\ref{sko1}) and (\ref{sko2})) that $E_{2}^{\prime }=\gamma E_{2}-\beta
\gamma cB_{3}$, see Jackson's book$^{(5)}$ Sec. 11.10. Thus in the usual
covariant approach the components $F^{a\beta }$ are transformed by the
passive LT into $F^{\prime a\beta }$ and then it is simply argued that six
independent components of $F^{\prime a\beta }$ are the ``Lorentz
transformed'' components $E_{i}^{\prime }$ and $B_{i}^{\prime }$. The
identification (\ref{sko2}) reveals an additional disadvantage in the usual
covariant approach that is not mentioned in Sec. 2.2.. It is

\noindent vi) It is not possible to speak about the LT of some components of
$F^{a\beta }$ as in Eq. (\ref{sko2}); \emph{the LT always transform the
whole geometric 4D quantity} and not some components. Further, by the same
procedure as in Sec. 2.2., one finds the ``transformed'' equations of the
same form as Eqs. (\ref{j3}) and (\ref{max}), but with primed quantities
replacing the unprimed ones, e.g., $\partial _{k}^{\prime }E_{k}^{\prime
}-j^{\prime 0}/c\varepsilon _{0}=0,$ and
\begin{equation}
\nabla ^{\prime }\mathbf{E}^{\prime }(\mathbf{r}^{\prime },t^{\prime })=\rho
^{\prime }(\mathbf{r}^{\prime },t^{\prime })/\varepsilon _{0},...,
\label{max1}
\end{equation}
where, e.g., the 3D vector $\mathbf{E}^{\prime }$ is again obtained
multiplying the components $F^{\prime i0}$ by \emph{the unit 3D vectors }$%
\mathbf{i}^{\prime },$ $\mathbf{j}^{\prime },$ $\mathbf{k}^{\prime }\mathbf{,%
}$ $\mathbf{E}^{\prime }\mathbf{=}F^{\prime 10}\mathbf{i}^{\prime
}+F^{\prime 20}\mathbf{j}^{\prime }+F^{\prime 30}\mathbf{k}^{\prime }$.
However \emph{the meaning of the 3D vectors}\textbf{\ }$\mathbf{i}^{\prime
}, $\textbf{\ }$\mathbf{j}^{\prime },$\textbf{\ }$\mathbf{k}^{\prime }$ is
undefined\textbf{;}\emph{\ they are not obtained by any transformation,
particularly not by the LT}\textbf{\ }\emph{from the 3D vectors}\textbf{\ }$%
\mathbf{i},$\textbf{\ }$\mathbf{j},$\textbf{\ }$\mathbf{k}$\textbf{. }%
Obviously such procedure has the same disadvantages as those discussed in
Sec. 2.2 including the new one, vi). The components $E_{i}^{\prime }$, $%
B_{i}^{\prime }$ and the 3D fields are all ill-defined in the 4D spacetime.
On the other hand the meaning of all quantities in the above geometric
approach is very clear; they are all well-defined in the 4D spacetime.
Moreover, the difference between the LT and the ST of the 4D quantities
representing the electric and magnetic fields is clearly seen; in the LT
always the whole 4D geometric quantity is transformed as, e.g., in Eqs. (\ref
{nle}) and (\ref{nlb}), whereas in the ST only a part of the whole 4D
geometric quantity is transformed as, e.g., in Eqs. (\ref{ce}) and (\ref{B}%
). Nevertheless the usual procedure, the identifications (\ref{sko1}) and (%
\ref{sko2}) and the derivation of the ``transformed'' equations (\ref{max1})
is considered for almost hundred years as relativistically correct
procedure. It is argued in every paper and textbook on the relativistic
electrodynamics (without exception as I am aware) that the equations (\ref
{max}) are Lorentz covariant equations, i.e., that the LT of the equations (%
\ref{max}) are the equations (\ref{max1}). Our discussion explicitly shows
that in the 4D spacetime the usual procedure is not justified either
mathematically or physically. \bigskip \medskip

\noindent \textbf{2.6. Lorentz\ Invariant\ Field\ Equations\ with 1-Vectors }%
$E,$\textbf{\ }$B$

\textbf{and Bivectors }$E_{Hv},$ $B_{Hv}$ \bigskip

Let us now remove the disadvantage mentioned at the end of Sec. 2.2. that
still exists in all Clifford algebra approaches to the electromagnetism.
Instead of decomposing $F$ into the \emph{observer dependent }$E_{f}$ and $%
B_{f}$ in the $\gamma _{0}$ - frame, as in Eq. (\ref{ebg}), we present the
decomposition of $F$ into the AQs, 1-vectors of the electric $E$ and
magnetic $B$ fields that are defined without reference frames, see also Ref.
15. We define
\begin{align}
F& =(1/c)E\wedge v+(IB)\cdot v,  \notag \\
E& =(1/c)F\cdot v,\quad IB=(1/c^{2})F\wedge v,\ B=-(1/c^{2})I(F\wedge v),
\label{myF}
\end{align}
where $I$ is the unit pseudoscalar. ($I$ is defined algebraically without
introducing any reference frame, as in Ref. 19 Sec. 1.2.) It holds that $%
E\cdot v=B\cdot v=0$ (since $F$ is skew-symmetric). $v$ in Eq. (\ref{myF})
can be interpreted as the velocity (1-vector) of a family of observers who
measures $E$ and $B$ fields. \emph{The velocity} $v$ \emph{and all other
quantities entering into} Eq. (\ref{myF}) \emph{are defined without
reference frames.} $v$ characterizes some general observer. Thus \emph{the
relations }(\ref{myF})\emph{\ hold for any observer.} However it has to be
emphasized that Eq. (\ref{myF}) is not a physical definition of $E$ and $B;$
the physical definition has to be given in terms of the Lorentz force and
Newton's second law as, e.g., in Ref. 15. The relations (\ref{myF}) actually
establish the equivalence of the formulation of electrodynamics with the
field bivector $F,$ see Ref. 10, and the formulation with 1-vectors of the
electric $E$ and magnetic $B$ fields. \emph{Both formulations, with} $F$
\emph{and} $E,$ $B$ \emph{fields, are equivalent formulations, but every of
them is a complete, consistent and self-contained formulation.} When Eq. (%
\ref{myF}) is inserted into the field equation for $F$, Eq. (\ref{MEF}),
then Eq. (\ref{MEF}) becomes the field equation for $E,$ $B$ fields
\begin{equation}
\partial \lbrack E\wedge (v/c)+(IB)\cdot v]=j/\varepsilon _{0}c.  \label{deb}
\end{equation}
In contrast to the field equation (\ref{eqfi}), that holds only for the $%
\gamma _{0}$-observer, the field equation (\ref{deb}) \emph{holds for any
observer;} \emph{the quantities entering into} Eq. (\ref{deb}) \emph{are all
AQs.} \emph{The equation} (\ref{deb}) \emph{is physicaly completely
equivalent to the field equation for} $F$ (\ref{MEF}). In some basis $%
\left\{ e_{\mu }\right\} $ the field equation (\ref{deb}) can be written as
a CBGE
\begin{align}
\lbrack \partial _{\alpha }(\delta _{\quad \mu \nu }^{\alpha \beta }E^{\mu
}v^{\nu }+\varepsilon ^{\alpha \beta \mu \nu }v_{\mu }cB_{\nu })-& (j^{\beta
}/\varepsilon _{0})]e_{\beta }+  \notag \\
\partial _{\alpha }(\delta _{\quad \mu \nu }^{\alpha \beta }v^{\mu }cB^{\nu
}+\varepsilon ^{\alpha \beta \mu \nu }v_{\mu }E_{\nu })e_{5}e_{\beta }& =0,
\label{maeb}
\end{align}
where $E^{\alpha }$ and $B^{\alpha }$ are the basis components of the
electric and magnetic 1-vectors $E$ and $B$, $\delta _{\quad \mu \nu
}^{\alpha \beta }=\delta _{\,\,\mu }^{\alpha }\delta _{\,\,\nu }^{\beta
}-\delta _{\,\,\nu }^{\alpha }\delta _{\,\mu }^{\beta }$ and $e_{5}$ is the
pseudoscalar for the frame $\{e_{\mu }\}$. The first part in Eq. (\ref{maeb}%
) (it contains sources) emerges from $\partial \cdot F=j/\varepsilon _{0}c$
and the second one (the source-free part) is obtained from $\partial \wedge
F=0,$ see also Ref. 15. Instead of working with the observer independent
field equation in the $F$- formulation, Eq. (\ref{MEF}), one can
equivalently use the $E,$ $B$ - formulation with the field equation (\ref
{deb}), or in the $\left\{ e_{\mu }\right\} $ basis Eq. (\ref{maeb}). (The
complete $E,$ $B$ formulation of relativistic electrodynamics will be
reported elsewhere.) Furthermore one can completely forget the manner in
which the equation with $E$ and $B$ is obtained, i.e., the field equation
with $F$ (\ref{MEF}), \emph{and consider the equation with} $E$ \emph{and} $%
B $, Eq. (\ref{deb}), \emph{which is defined without reference frames, or
the corresponding CBGE} (\ref{maeb}), \emph{as the primary and fundamental
equations for the whole classical electromagnetism. }In such a correct
relativistic formulation of electromagnetism the field equation with 1-
vectors $E$ and $B$, Eq. (\ref{deb}), takes over the role of the usual ME
with the 3D $\mathbf{E}$ and $\mathbf{B}$, i.e., of the ME (\ref{MEC}). We
note that the equivalent formulation of electrodynamics with tensors $E^{a}$
and $B^{a}$ is reported in Refs. 11, 20, whereas the component form in the
Einstein system of coordinates\ is given in Refs. 17,21 and Ref. 22.

Let us now take that in Eq. (\ref{maeb}) the standard basis $\left\{ \gamma
_{\mu }\right\} $ is used instead of some general basis $\left\{ e_{\mu
}\right\} .$ Then Eq. (\ref{maeb}) can be written as $C^{\beta }\gamma
_{\beta }+D^{\beta }\gamma _{5}\gamma _{\beta }=0,$ where $C^{\beta
}=\partial _{\alpha }(\delta _{\quad \nu \mu }^{\alpha \beta }v^{\mu }E^{\nu
}-\varepsilon ^{\alpha \beta \nu \mu }v_{\mu }cB_{\nu })-j^{\beta
}/\varepsilon _{0}$ and $D^{\beta }=\partial _{\alpha }(\delta _{\quad \mu
\nu }^{\alpha \beta }v^{\mu }cB^{\nu }+\varepsilon ^{\alpha \beta \mu \nu
}v_{\mu }E_{\nu })$. \emph{When the active LT are applied to} \emph{Eq. }(%
\ref{maeb}) \emph{with the} $\left\{ \gamma _{\mu }\right\} $ \emph{basis
the equation remains of the same form but with primed quantities replacing
the unprimed ones (of course the basis is unchanged).} This can be
immediately seen since the equation (\ref{maeb}) is written in a manifestly
covariant form. Thus the Lorentz transformed Eq. (\ref{maeb}) is
\begin{align}
R(C^{\beta }\gamma _{\beta }+D^{\beta }\gamma _{5}\gamma _{\beta })%
\widetilde{R}& =0,  \notag \\
C^{\prime \beta }\gamma _{\beta }+D^{\prime \beta }\gamma _{5}\gamma _{\beta
}& =0,  \label{clo}
\end{align}
where, e.g., $C^{\prime \beta }=\partial _{\alpha }^{\prime }(\delta _{\quad
\nu \mu }^{\alpha \beta }v^{\prime \mu }E^{\prime \nu }-\varepsilon ^{\alpha
\beta \nu \mu }v_{\mu }^{\prime }cB_{\nu }^{\prime })-j^{\prime \beta
}/\varepsilon _{0}$. Obviously \emph{such a formulation of electromagnetism
with the fundamental equation} (\ref{deb}) \emph{or} (\ref{maeb}) \emph{is a
relativistically correct formulation.}

What is the relation between the relativistically correct field equation (%
\ref{deb}) or (\ref{maeb}) and the usual ME (\ref{MEC})? From the above
discussion and from Sec. 2.1. one concludes that if in Eq. (\ref{deb}) we
specify the velocity $v$ of the observers who measure $E$ and $B$\ fields to
be $v=c\gamma _{0}$, then the equation (\ref{deb}) becomes the equation (\ref
{eqfi}). Further choosing the standard basis $\left\{ \gamma _{\mu }\right\}
$ in the $\gamma _{0}$ - frame, in which $v=c\gamma _{0}$, or in the
components $v^{\alpha }=(c,0,0,0)$, then in that $\gamma _{0}$ - frame $E$
and $B$ become $E_{f}$ and $B_{f}$ and they do not have temporal components,
$E_{f}^{0}=B_{f}^{0}=0$. \emph{The CBGE} (\ref{maeb}) \emph{becomes the
usual ME} (\ref{MEC}). Thus the usual Clifford algebra treatments of
electromagnetism$^{(7,8,9)}$ with the space-time split in the $\gamma _{0}$
- frame and the usual ME (\ref{MEC}) are simply obtained from our \emph{%
observer independent} formulation with field equation (\ref{deb}) or (\ref
{maeb}) choosing that $v=c\gamma _{0}$ and choosing the standard basis $%
\left\{ \gamma _{\mu }\right\} $. We see that the \emph{correspondence
principle} is simply satisfied in this formulation with $E$ and $B$\ fields;
all results obtained in the previous treatments from the usual ME with the
3D $\mathbf{E}$ and $\mathbf{B}$ remain valid in the formulation with the
1-vectors $E$ and $B$ if physical phenomena are considered only in one
inertial frame of reference. Namely the selected inertial frame of reference
can be chosen to be the $\gamma _{0}$ - frame with the $\left\{ \gamma _{\mu
}\right\} $ basis. Then there, as explained above, the CBGE (\ref{maeb}) can
be reduced to the equations containing only the components, the four ME in
the component form, the ME (\ref{MEC}). Thus for observers who are at rest
in the $\gamma _{0}$ - frame ($v=c\gamma _{0}$) the components of the 3D $%
\mathbf{E}$ and $\mathbf{B}$ can be simply replaced by the space components
of the 1-vectors $E$ and $B$ in the $\left\{ \gamma _{\mu }\right\} $ basis.
We remark that just such observers are usually considered in the
conventional formulation with the 3D $\mathbf{E}$ and $\mathbf{B.}$ The
dependence of the field equations (\ref{maeb}) on $v$ reflects the
arbitrariness in the selection of the $\gamma _{0}$ - frame but at the same
time it makes the equations (\ref{maeb}) independent of that choice. The $%
\gamma _{0}$ - frame can be selected at our disposal, which proves that we
don't have a kind of the ``preferred'' frame theory. \emph{All experimental
results that are obtained in one inertial frame of reference can be equally
well explained by our geometric formulation of the electromagnetism with the
1-vectors} $E$ \emph{and} $B$ \emph{as they are explained by the usual ME
with the 3D} $\mathbf{E}$ \emph{and }$\mathbf{B.}$

\emph{However there is a fundamental difference between the standard
approach with the 3D} $\mathbf{E}$ \emph{and} $\mathbf{B}$ \emph{and the
approach with the 4D AQs }$E$ \emph{and} $B$. It is considered in all
standard treatments that the equation (\ref{EBC}) is the LT\ of the original
ME (\ref{MEC}). But, as shown here, the equation (\ref{EBC}) is not the LT
of the original ME (\ref{MEC}); the LT of the ME (\ref{MEC}) are the
equations (\ref{L}) (i.e., (\ref{L1}) with (\ref{anu}), or (\ref{mcr})). The
ME (\ref{MEC}) are obtained from our field equation (\ref{maeb}) putting $%
v=c\gamma _{0}$ and choosing the standard basis $\left\{ \gamma _{\mu
}\right\} $. In the same way the equations (\ref{clo}), which are the LT of
the equations (\ref{maeb}), become the LT of the ME (\ref{MEC}), that is,
the equations (\ref{L}) (or (\ref{L1}) with (\ref{anu}), or (\ref{mcr})),
when in Eq. (\ref{clo}) it is taken that $v^{\prime }$, $\partial ^{\prime }$%
, $E^{\prime }$ and $B^{\prime }$ are the LT of $v=c\gamma _{0}$, $\partial $%
, $E_{f}$ and $B_{f}$, that is, $v^{\prime }=R(c\gamma _{0})\widetilde{R}$, $%
\partial ^{\prime }=R\partial \widetilde{R}$, $E^{\prime }=RE_{f}\widetilde{R%
}=E_{f}^{\prime },$ $B^{\prime }=RB_{f}\widetilde{R}=B_{f}^{\prime }$. We
recall from Sec. 2.3. that to an observer in the $\left\{ \gamma _{\mu
}\right\} $ frame the vector $p^{\prime }$ ($p^{\prime }=Rp\widetilde{R}%
=p^{\prime \mu }\gamma _{\mu }$) appears the same as the vector $p$ ($%
p=p^{\mu }\gamma _{\mu }$) appears to an observer in the $\left\{ \gamma
_{\mu }^{\prime }\right\} $ frame. This, together with the preceding
discussion, show that \emph{the usual ME with the 3D} $\mathbf{E}$ \emph{and}
$\mathbf{B}$, \emph{i.e., the equation} (\ref{MEC}) \emph{and the equation} (%
\ref{EBC}) \emph{obtained by the ST from} (\ref{MEC}), \emph{cannot be used
for the explanation of any experiment that tests SR}, i.e., in which
relatively moving observers have to compare their data \emph{obtained by
measurements on the same physical object.} In contrast to the description of
the electromagnetism with the 3D $\mathbf{E}$ and $\mathbf{B,}$ \emph{the
description with the 4D fields }$E$ \emph{and} $B$, \emph{i.e., with the
equations }(\ref{maeb}) \emph{and} (\ref{clo}), \emph{is correct not only in
the} $\gamma _{0}$ - \emph{frame with the standard basis} $\left\{ \gamma
_{\mu }\right\} $ \emph{but in all other relatively moving frames and it
holds for any permissible choice of coordinates, i.e., basis} $\left\{
e_{\mu }\right\} $. We see that the relativistically correct fields $E$ and $%
B$ and the new field equations (\ref{deb}) and (\ref{maeb}) do not have the
same physical interpretation as the usual 3D fields $\mathbf{E}$ and $%
\mathbf{B}$ and the usual 3D ME (\ref{MEC}) except in the $\gamma _{0}$ -
frame with the $\left\{ \gamma _{\mu }\right\} $ basis in which $%
E^{0}=B^{0}=0$. This consideration completely defines the relation between
our approach with 4D $E$ and $B$ and all previous approaches.

As explained in the preceding sections the \emph{observer independent} $F$
field is decomposed in Eq. (\ref{FB}), see Refs. 7, 8, in terms of \emph{%
observer dependent quantities,} i.e., as the sum of a relative vector $%
\mathbf{E}_{H}$ and a relative bivector $\gamma _{5}\mathbf{B}_{H},$ by
making the space-time split in the $\gamma _{0}$ - frame. But, here we
present the\emph{\ new }decomposition of $F$ into the AQs, the bivectors $%
E_{Hv}$ and $B_{Hv}$, \emph{which are independent of the chosen reference
frame and of the chosen system of coordinates in it.} We define
\begin{align}
F& =E_{Hv}+cIB_{Hv}\mathbf{,\quad }E_{Hv}=(1/c^{2})(F\cdot v)\wedge v  \notag
\\
B_{Hv}& =-(1/c^{3})I[(F\wedge v)\cdot v],\quad IB_{Hv}=(1/c^{3})(F\wedge
v)\cdot v  \label{he}
\end{align}
(The subscript $Hv$ is for ``Hestenes'' with $v$ and not, as usual, Refs.
7,8, with $\gamma _{0}$.) Obviously Eq. (\ref{he}) \emph{holds for any
observer.} When we use Eq. (\ref{he}) in the field equation for $F$ (\ref
{MEF}), and after multiplication by $v/c$ (instead of by $\gamma _{0}$), the
equation (\ref{MEF}) becomes
\begin{equation}
(v/c)\{\partial (E_{Hv}+cIB_{Hv})-j/\varepsilon _{0}c\}=0.  \label{Nf}
\end{equation}
In contrast to the field equation (\ref{H1}) that holds only for the $\gamma
_{0}$-observer, the field equation (\ref{Nf}) \emph{holds for any observer;}
\emph{the quantities entering into} Eq. (\ref{Nf}) \emph{are the AQs.} \emph{%
The equation} (\ref{Nf}) \emph{is physically completely equivalent to the
field equation for} $F$ (\ref{MEF})\emph{, i.e.,} \emph{to the field
equation with 1- vectors} $E$ \emph{and} $B$ (\ref{deb}). (The equation (\ref
{H1}) corresponds to the equation (\ref{eqfi}), whereas Eq. (\ref{Nf})
corresponds to Eq. (\ref{deb}).) The field equation (\ref{Nf}) can be
written as a CBGE, and it looks much more complicated than the equation (\ref
{maeb}) with 1- vectors $E$ and $B$. We write it (for better comparison) as
two equations; the first one will yield the scalar and bivector parts of Eq.
(\ref{H2}) when $v/c=\gamma _{0}$. It is
\begin{align}
& (1/c)v_{\beta }\partial _{\alpha }(E_{Hv})^{\alpha \beta
}+[(1/2c)v^{\alpha }\partial _{\alpha }(E_{Hv})^{\beta \sigma
}-(1/2)\varepsilon ^{\mu \nu \alpha \sigma }v^{\beta }\partial _{\alpha
}(B_{Hv})_{\mu \nu }]\gamma _{\beta }\wedge \gamma _{\sigma }  \notag \\
& =(1/\varepsilon _{0}c^{2})(v_{\alpha }j^{\alpha }+v^{\beta }j^{\sigma
}\gamma _{\beta }\wedge \gamma _{\sigma }).  \label{nh}
\end{align}
The second equation will yield the pseudoscalar and pseudobivector parts of
Eq. (\ref{H2}) when $v/c=\gamma _{0}$ and it is
\begin{equation}
v_{\beta }\partial _{\alpha }(B_{Hv})^{\alpha \beta }\gamma
_{5}+(1/2)v^{\alpha }\partial _{\alpha }(B_{Hv})^{\mu \nu }\gamma
_{5}(\gamma _{\mu }\wedge \gamma _{\nu })+(v_{\beta }\partial ^{\alpha
}-v^{\alpha }\partial _{\beta })(E_{Hv})_{\alpha \nu }\gamma ^{\beta }\wedge
\gamma ^{\nu }=0.  \label{D}
\end{equation}
In the $\left\{ \gamma _{\mu }\right\} $ basis $I=\gamma _{5}$. The equation
(\ref{nh}) is with sources and it emerges from $\partial \cdot
F=j/\varepsilon _{0}c$, while Eq. (\ref{D}) is the source-free equation and
it emerges from $\partial \wedge F=0$. Comparing Eqs. (\ref{nh}) and (\ref{D}%
) in the $E_{Hv}$, $B_{Hv}$ - formulation with the corresponding parts in
Eq. (\ref{maeb}) with 1- vectors $E$ and $B$ we see that the formulation
with $E$ and $B$ is much simpler and more elegant than the formulation with
bivectors $E_{Hv}$ and $B_{Hv}$; the physical content is completely
equivalent.

The equations (\ref{nh}) and (\ref{D}) are written in a manifestly covariant
form. This means that \emph{when the active LT are applied upon such} \emph{%
Eqs.} (\ref{nh}) \emph{and} (\ref{D}) \emph{the equations remain of the same
form but with primed quantities replacing the unprimed ones (of course the
basis is unchanged). }

The whole discussion with 1- vectors $E$ and $B$ about the correspondence
principle applies in the same measure to the formulation with bivectors $%
E_{Hv}$ and $B_{Hv}$. The only difference is the simplicity of the
formulation with 1- vectors $E$ and $B$.

The same conclusions hold for the formulation with 1-vector $\mathbf{E}_{J}$
and a bivector $\mathbf{B}_{J}$ from Ref. 9, but for the sake of brevity
that formulation will not be considered here. \bigskip \medskip

\noindent \textbf{3. THE PROOF OF THE DIFFERENCE\ BETWEEN\ THE\ ST }

\textbf{AND THE\ LT\ OF\ THE ME\ USING\ THE TENSOR }

\textbf{FORMALISM WITH\ 4-VECTORS\ }$E^{a}$\textbf{\ AND\ }$B^{a}$\bigskip

The same proof and the whole consideration as with 1-vectors $E$ and $B$ can
be given in the tensor formalism as well (it is presented in detail in Ref.
13). The important parts of this issue are already treated in Refs. 11, 15,
1. Therefore we only quote the main results. Now we start with Lorentz
invariant field equations with $v$ and with the decomposition of $F^{ab}$
into the AQs $E^{a}$ and $B^{a}$ since in the tensor formalism such field
equations and the decomposition are already in use, Refs. 23, 24.

The electromagnetic field tensor $F^{ab}$ is defined as an AQ; it is an
abstract tensor. Latin indices a,b,c, ... are to be read according to the
abstract index notation, as in Refs. 23, 24 and Refs. 11, 12, 20. As already
said in the invariant SR that uses 4D AQs in the tensor formalism, Refs. 11,
12, 20, 1, and in the Clifford algebra formalism, Refs. 10, 15, 2, any
permissible system of coordinates, not necessary the Einstein system of
coordinates, i.e., the standard basis $\left\{ \gamma _{\mu }\right\} $, can
be used on an equal footing. However, for simplicity, we shall only deal
with the standard basis $\left\{ \gamma _{\mu }\right\} $. In the tensor
formalism $\gamma _{\mu }$ denote the basis 4-vectors forming the standard
basis $\left\{ \gamma _{\mu }\right\} $.

In the abstract index notation the field equations with $F^{ab}$ are given
as
\begin{equation}
(-g)^{-1/2}\partial _{a}((-g)^{1/2}F^{ab})=j^{b}/\varepsilon _{0}c,\quad
\varepsilon ^{abcd}\partial _{b}F_{cd}=0,  \label{maxten}
\end{equation}
where $g$ is the determinant of the metric tensor $g_{ab}$ and $\partial
_{a} $ is an ordinary derivative operator. Now there are two field equations
whereas in the geometric algebra formalism they are united in only one field
equation. When written in the $\left\{ \gamma _{\mu }\right\} $ basis as
CBGEs the relations (\ref{maxten}) become
\begin{equation}
\partial _{\alpha }F^{a\beta }\gamma _{\beta }=(1/\varepsilon _{0}c)j^{\beta
}\gamma _{\beta },\quad \partial _{\alpha }\ ^{\ast }F^{\alpha \beta }\gamma
_{\beta }=0.  \label{maco1}
\end{equation}
Instead of Eq. (\ref{myF}) from Sec. 2.6. we have the decomposition of $%
F^{ab}$ into the AQs, the 4-vectors $E^{a}$ and $B^{a}$
\begin{align}
F^{ab}& =(1/c)\delta _{\quad cd}^{ab}E^{c}v^{d}+\varepsilon
^{abcd}v_{c}B_{d},  \notag \\
E^{a}& =(1/c)F^{ab}v_{b},\quad B^{a}=(1/2c^{2})\varepsilon
^{abcd}v_{b}F_{cd}.  \label{cf}
\end{align}
Inserting Eq. (\ref{cf}) into Eq. (\ref{maxten}) we find the Lorentz
invariant field equations with $E^{a}$ and $B^{a}$ that correspond to Eq. (%
\ref{deb}) from Sec. 2.6. When these equations are written as CBGEs in the $%
\left\{ \gamma _{\mu }\right\} $ basis they become
\begin{align}
\partial _{\alpha }(\delta _{\quad \mu \nu }^{\alpha \beta }E^{\mu }v^{\nu
}+\varepsilon ^{\alpha \beta \mu \nu }v_{\mu }cB_{\nu })\gamma _{\beta }&
=(j^{\beta }/\varepsilon _{0})\gamma _{\beta }  \notag \\
\partial _{\alpha }(\delta _{\quad \mu \nu }^{\alpha \beta }v^{\mu }cB^{\nu
}+\varepsilon ^{\alpha \beta \mu \nu }v_{\mu }E_{\nu })\gamma _{\beta }& =0.
\label{em2}
\end{align}
The equations (\ref{em2}) correspond to Eq. (\ref{maeb}) from Sec. 2.6.
(when written in the standard basis $\left\{ \gamma _{\mu }\right\} )$. It
is clear from the form of \emph{the equations} (\ref{em2})\emph{\ (with some
general }$v^{\mu }$\emph{)} \emph{that they} \emph{are invariant upon the
passive LT}$.$ Namely in a relatively moving frame $S^{\prime }$ all
quantities in (\ref{em2}) will be replaced with the primed quantities that
are obtained by the passive LT (of course, $\delta _{\quad \mu \nu }^{\alpha
\beta }$ and $\varepsilon ^{\alpha \beta \mu \nu }$ are unchanged). \emph{%
All the primed quantities (components and the basis) are obtained from the
corresponding unprimed quantities through the LT.} The components of any 4D
CBGQ\ transform by the LT, while the basis vectors $\gamma _{\mu }$
transform by the inverse LT, \emph{thus leaving the whole 4D CBGQ} \emph{%
invariant upon the passive LT. }The invariance of some 4D CBGQ\ upon the
passive LT reflects the fact that such 4D quantity represents \emph{the same
physical object} for relatively moving observers. Due to the invariance of
every CBGQ upon the passive LT the field equations with primed quantities,
thus in $S^{\prime }$, are \emph{exactly equal} to the corresponding
equations in $S,$ given by Eq. (\ref{em2}). Thus the equations (\ref{em2})
are not only covariant but also the Lorentz invariant field equations. The
principle of relativity is automatically included in such formulation.

The usual ME are simply obtained from eq. (\ref{em2}) specifying that $%
v^{\alpha }=c(\gamma _{0})^{\alpha }$, i.e., choosing the rest frame of
``fiducial'' observers, the $\gamma _{0}$ - frame with the $\left\{ \gamma
_{\mu }\right\} $ basis. Then from Eq. (\ref{em2}) we first find the ME
exactly corresponding to Eq. (\ref{cl}) from Sec. 2.1. and further the
component form of the usual ME corresponding to Eq. (\ref{MEC}) (but now
there are two equations)
\begin{align}
(\partial _{k}E_{f}^{k}-j^{0}/c\varepsilon _{0})\gamma _{0}+(-\partial
_{0}E_{f}^{i}+c\varepsilon ^{ijk0}\partial _{j}B_{fk}-j^{i}/c\varepsilon
_{0})\gamma _{i}& =0  \notag \\
(-c\partial _{k}B_{f}^{k})\gamma _{0}+(c\partial _{0}B_{f}^{i}+\varepsilon
^{ijk0}\partial _{j}E_{fk})\gamma _{i}& =0.  \label{me}
\end{align}
As in Sec. 2.1., in the $\gamma _{0}$ - frame with the $\left\{ \gamma _{\mu
}\right\} $ basis, $E_{f}^{0}=B_{f}^{0}=0$, and the relations (\ref{sko})
hold also here $E_{f}^{i}=F^{i0}$, $B_{f}^{i}=(-1/2c)\varepsilon
^{0kli}F_{kl}$ (the standard identification), since $E_{f}^{\mu }=F^{\mu \nu
}(\gamma _{0})_{\nu }$, $B_{f}^{\mu }=(1/c)(F^{\ast })^{\mu \nu }(\gamma
_{0})_{\nu }$. The equations (\ref{me}) (and (\ref{em2}) as well) can be
written as $a^{\alpha }\gamma _{\alpha }=0$ and $b^{\alpha }\gamma _{\alpha
}=0.$ The coefficients $a^{\alpha }$ and $b^{\alpha }$ are clear from the
first and second equation respectively in Eq. (\ref{me}); \emph{they are the
usual ME in the component form}.

Let us now apply the passive LT to the ME (\ref{me}). Upon the passive LT
the sets of components $E_{f}^{\mu }$ and $B_{f}^{\mu }$ and the basis $%
\left\{ \gamma _{\mu }\right\} $ of the $\gamma _{0}$ - frame (the $S$
frame) transform to $E_{f}^{\prime \mu }$ and $B_{f}^{\prime \mu }$ and the
new basis $\left\{ \gamma _{\mu }^{\prime }\right\} $ in the relatively
moving inertial frame of reference $S^{\prime }$, e.g., $E^{\prime \nu
}=L_{\ \delta }^{\nu }E^{\delta }$ and $\gamma _{\mu }^{\prime }=(L^{-1})_{\
\mu }^{\delta }\gamma _{\delta }$ (the components $L_{\ \delta }^{\nu }$ are
quoted in Sec. 2.5). For the boost in the $\gamma _{1}$ direction the
Lorentz transformed sets of components $E_{f}^{\prime \mu }$ and $%
B_{f}^{\prime \mu }$ are given as
\begin{align}
E_{f}^{\prime \mu }& =(1/c)F^{\prime \mu \nu }v_{\nu }^{\prime },\quad
E_{f}^{\prime \mu }=\left( -\beta \gamma E^{1},\gamma
E^{1},E^{2},E^{3}\right) ,  \notag \\
B_{f}^{\prime \mu }& =(1/c^{2})(F^{\ast })^{\prime \mu \nu }v_{\nu }^{\prime
},\quad B_{f}^{\prime \mu }=\left( -\beta \gamma B^{1},\gamma
B^{1},B^{2},B^{3}\right) ,  \label{ebcr}
\end{align}
where $v_{\nu }^{\prime }=\left( c\gamma ,c\beta \gamma ,0,0\right) ,$ and $%
v^{\prime \nu }/c$, as the LT of $(\gamma _{0})^{\nu }$, $v^{\prime \nu
}/c=L_{\ \delta }^{\nu }(\gamma _{0})^{\delta }$, is not in the time
direction in $S^{\prime }$, i.e., it is not $=(\gamma _{0}^{\prime })^{\nu }$%
. Note that $E_{f}^{\prime \mu }$\emph{\ and }$B_{f}^{\prime \mu }$\emph{\
have the temporal components as well. }Further \emph{the components }$%
E_{f}^{\mu }$ ($B_{f}^{\mu }$) \emph{in }$S$ \emph{transform upon the LT
again to the components} $E_{f}^{\prime \mu }$ ($B_{f}^{\prime \mu }$) \emph{%
in} $S^{\prime }$\emph{; there is no mixing of components.} Actually this is
the way in which every well-defined 4-vector (the components) transforms
upon the LT. The relations (\ref{ebcr}) are given in Ref. 1 and they
correspond to relations (\ref{nle}) and (\ref{nlb}) from Sec. 2.3.. The AQ,
e.g., an abstract tensor $E^{a},$ can be represented by CBGQs in $S$ and $%
S^{\prime }$ as $E_{f}^{\mu }\gamma _{\mu }$ and $E_{f}^{\prime \mu }\gamma
_{\mu }^{\prime }$ and, of course, it must hold that, e.g., $%
E^{a}=E_{f}^{\mu }\gamma _{\mu }=E_{f}^{\prime \mu }\gamma _{\mu }^{\prime }$%
. Then the equations (\ref{me}) transform to
\begin{equation}
a^{\prime \alpha }\gamma _{\alpha }^{\prime }=0,\quad b^{\prime \alpha
}\gamma _{\alpha }^{\prime }=0,  \label{abc}
\end{equation}
and it holds, as for any 4-vector (a geometric quantity), that $a^{\prime
\alpha }\gamma _{\alpha }^{\prime }=a^{\alpha }\gamma _{\alpha },$ and $%
b^{\prime \alpha }\gamma _{\alpha }^{\prime }=b^{\alpha }\gamma _{\alpha }$;
the coefficients transform by the LT, e.g. $a^{\prime 0}=\gamma a^{0}-\beta
\gamma a^{1}$, while the basis 4-vectors transform by the inverse LT, e.g., $%
\gamma _{0}^{\prime }=\gamma \gamma _{0}+\beta \gamma \gamma _{1}$. Of
course $(\gamma _{0})^{\nu }$ transforms to $v^{\prime \nu }/c$ and $%
E_{f}^{\prime \mu }$, $B_{f}^{\prime \mu }$ are given by Eq. (\ref{ebcr}).
When the coefficients $a^{\prime \alpha }$ and $b^{\prime \alpha }$ are
written in terms of the primed quantities (from the $S^{\prime }$ frame)
then we find the same expressions as in Sec. 2.3., e.g., the expression (\ref
{anu}) is obtained for $a^{\prime 0}$, and, of course, $a^{\prime 0}$ \emph{%
is completely different in form than the coefficient }$a^{0}=(\partial
_{k}E_{f}^{k}-j^{0}/c\varepsilon _{0})$ in Eq. (\ref{me}). Thus these
Lorentz transformed ME exactly correspond to the equation (\ref{L1}) with
Eq. (\ref{anu}) from Sec. 2.3.. We again see that the usual ME are not
Lorentz covariant equations.

As shown above upon the LT $(\gamma _{0})^{\nu }$ transforms to $v^{\prime
\nu }/c=L_{\ \delta }^{\nu }(\gamma _{0})^{\delta }$, which is not $=(\gamma
_{0}^{\prime })^{\nu }$, i.e., it is not in the time direction in $S^{\prime
}$. However it is implicitly assumed in all usual treatments, e.g., Ref. 5
and Ref. 6 eqs. (3.5) and (3.24), that in $S^{\prime }$ one can again make
the identification of six independent components of $F^{\prime \mu \nu }$
with three components $E_{i}^{\prime }$, $E_{i}^{\prime }=F^{\prime i0}$,
and three components $B_{i}^{\prime }$, $B_{i}^{\prime }=(1/2c)\varepsilon
_{ikl}F_{lk}^{\prime }$, Eq. (\ref{sko2}), see Secs. 2.2. and 2.5. This
means that standard treatments assume that upon the passive LT the set of
components $(\gamma _{0})^{\nu }=\left( 1,0,0,0\right) $ from $S$ transforms
to $(\gamma _{0}^{\prime })^{\nu }=\left( 1,0,0,0\right) $ ($(\gamma
_{0}^{\prime })^{\nu }$ are the components of the unit 4-vector \emph{in the
time direction in }$S^{\prime }$), and consequently that, as shown in Ref.
1, $E_{f}^{\mu }$ and $B_{f}^{\mu }$ transform to $E_{st.}^{\prime \mu }$
and $B_{st.}^{\prime \mu }$ in $S^{\prime }$ as
\begin{align}
E_{st.}^{\prime \mu }& =F^{\prime \mu \nu }(\gamma _{0}^{\prime })_{\nu },\
E_{st.}^{\prime \mu }=\left( 0,E^{1},\gamma E^{2}-\gamma \beta B^{3},\gamma
E^{3}+\gamma \beta B^{2}\right) ,  \notag \\
B_{st.}^{\prime \mu }& =(1/c)(^{\ast }F)^{\prime \mu \nu }(\gamma
_{0}^{\prime })_{\nu },\ B_{st.}^{\prime \mu }=\left( 0,B^{1},\gamma
B^{2}+\gamma \beta E^{3},\gamma B^{3}-\gamma \beta E^{2}\right) .  \label{kr}
\end{align}
\emph{The temporal components of }$E_{st.}^{\prime \mu }$\emph{\ and }$%
B_{st.}^{\prime \mu }$\emph{\ in }$S^{\prime }$\emph{\ are again zero as are
the temporal components of }$E_{f}^{\mu }$\emph{\ and }$B_{f}^{\mu }$\emph{\
in }$S.$\emph{\ }This fact clearly shows that \emph{the transformations }(%
\ref{kr})\emph{\ are not the LT of some well-defined 4D quantities; the LT
cannot transform the unit 4-vector in the time direction in one frame} $S$
\emph{to the unit 4-vector in the time direction in another relatively
moving frame }$S^{\prime }.$ Obviously $E_{st.}^{\prime \mu }$ and $%
B_{st.}^{\prime \mu }$ are completely different quantities than $%
E_{f}^{\prime \mu }$ and $B_{f}^{\prime \mu }$, Eq. (\ref{ebcr}), that are
obtained by the correct LT. We can easily check that $E_{st.}^{\prime \mu
}\gamma _{\mu }^{\prime }\neq E_{f}^{\mu }\gamma _{\mu },$ and $%
B_{st.}^{\prime \mu }\gamma _{\mu }^{\prime }\neq B_{f}^{\mu }\gamma _{\mu
}. $ This means that, e.g., $E_{f}^{\mu }\gamma _{\mu }$ and $%
E_{st.}^{\prime \mu }\gamma _{\mu }^{\prime }$ \emph{are not the same
quantity for observers in} $S$ \emph{and} $S^{\prime }.$ As far as
relativity is concerned the quantities, e.g., $E_{f}^{\mu }\gamma _{\mu }$
and $E_{st.}^{\prime \mu }\gamma _{\mu }^{\prime },$ are not related to one
another. The observers in $S$ and $S^{\prime }$ are not looking at the same
physical object but at two different objects; \emph{every observer makes
measurement on its own object and such measurements are not related by the
LT.} From \emph{the relativistically incorrect transformations} (\ref{kr})
one simply derives the transformations of the spatial components $%
E_{st.}^{\prime i}$ and $B_{st.}^{\prime i}$, the relations (\ref{sk1}) from
Sec. 2.4., \emph{which are exactly the ST of components of the 3D} $\mathbf{E%
}$ \emph{and} $\mathbf{B}$. \emph{According to the ST} \emph{the transformed
components} $E_{st}^{\prime i},$ and $B_{st}^{\prime i},$ \emph{are
expressed by the mixture of components} $E_{f}^{i}$ \emph{and} $B_{f}^{i}.$
This completely differs from the correct LT (\ref{ebcr}). The
transformations (\ref{kr}) and the transformations for $E_{st.}^{\prime i}$
and $B_{st.}^{\prime i}$ (\ref{sk1}) are typical examples of the
``apparent'' transformations that are first discussed in Refs. 25 and 26.
The ``apparent'' transformations of the spatial distances (the Lorentz
contraction) and the temporal distances (the dilatation of time) are
elaborated in detail in Refs. 11, 12, see also Ref. 22. It is explicitly
shown in Ref. 12 that the true agreement with experiments that test SR
exists when the theory deals with well-defined 4D quantities, i.e., the
quantities that are invariant upon the passive LT. However new experiments
that test SR are continuosly published in leading physical journals, e.g.,
Ref. 27, and in these papers the dilatation of time and the Lorentz
contraction are still considered as fundamental relativistic effects. (These
experiments will be discussed in detail elsewhere.)

Let us now perform the ST of the ME (\ref{me}) supposing that $E_{f}^{\mu }$%
\ and $B_{f}^{\mu }$\ in $S$\ are transformed into $E_{st.}^{\prime \mu }$
and $B_{st.}^{\prime \mu }$ in $S^{\prime }$ according to Eq. (\ref{kr}).
They are
\begin{align}
(\partial _{k}^{\prime }E_{st}^{\prime k}-j^{\prime 0}/c\varepsilon
_{0})\gamma _{0}^{\prime }+(-\partial _{0}^{\prime }E_{st}^{\prime
i}+c\varepsilon ^{ijk0}\partial _{j}^{\prime }B_{st,k}^{\prime }-j^{\prime
i}/c\varepsilon _{0})\gamma _{i}^{\prime }& =0,  \notag \\
(-c\partial _{k}^{\prime }B_{st}^{\prime k})\gamma _{0}^{\prime }+(c\partial
_{0}^{\prime }B_{st}^{\prime i}+\varepsilon ^{ijk0}\partial _{j}^{\prime
}E_{st,k}^{\prime })\gamma _{i}^{\prime }& =0.  \label{cm}
\end{align}
These equations are of the same form as the original ME (\ref{me}), but $%
E_{f}^{\mu }$\ and $B_{f}^{\mu }$\ from $S$ are not transformed by the LT
than by the ST (\ref{kr}) into $E_{st.}^{\prime \mu }$ and $B_{st.}^{\prime
\mu }$ in $S^{\prime }.$ Thence \emph{the equations }(\ref{cm})\emph{\ are
not the correct LT, but relativistically incorrect transformations of the
original ME} (\ref{me}); \emph{the LT of the ME} (\ref{me}) \emph{are the
equations} (\ref{abc}) \emph{with} $a^{\prime 0}$ as in Eq. (\ref{anu}),
where the Lorentz transformed $E_{f}^{\prime \mu }$ and $B_{f}^{\prime \mu }$
are given by the relations (\ref{ebcr}).\bigskip \medskip

\noindent \textbf{4.} \textbf{SHORT COMPARISON WITH EXPERIMENTS. }

\textbf{FARADAY DISK\bigskip }

Let us now briefly discuss, as an example, the Faraday disk, using both the
conventional formulation of electromagnetism with the 3D $\mathbf{E}$ and $%
\mathbf{B}$ and their ST and this new formulation, the invariant
relativistic electrodynamics, with geometric 4D quantities. The comparison
will be made in the tensor formalism from Sec. 3. since it is better known
for physicists than the geometric algebra formalism. A conducting disk is
turning about a thin axle passing through the center at a right angle to the
disk and parallel to a uniform magnetic field $\mathbf{B}$. The circuit is
made by connecting one end of the resistor to the axle (the spatial point $A$%
) and the other end to a sliding contact touching the external circumference
(the spatial point $C$). The disk of radius $R$ is rotating with angular
velocity $\omega .$ (For the description and the picture of the Faraday disk
see, e.g., Ref. 28 Chap. 18 or the recent paper.$^{(29)}$) Let us determine
the electromotive force (emf) in two inertial frames of reference, the
laboratory frame $S$ in which the disk is rotating and the frame $S^{\prime
} $ instantaneously co-moving with a point on the external circumference
(say $C$, taken at some moment $t$, e.g., $t=0$). The $x^{\prime }$ axis is
along the 3-velocity $\mathbf{V}$ of the point $C$ at $t$ and it is parallel
to the $x$ axis. Actually all axes in $S^{\prime }$ are parallel to the
corresponding axes in $S$. The $y^{\prime }$ axis is along the radius, i.e.,
along the segment $AC.$

First we calculate the emf using the standard formulation. In the $S$ frame
\begin{equation}
emf=\oint (\mathbf{F}_{L}/q)\cdot \mathbf{dl=}\int_{AC}(F_{L,y}/q)dy\mathbf{=%
}\omega R^{2}B/2,  \label{emf1}
\end{equation}
where $\mathbf{F}_{L}$ is the usual form for the 3D Lorentz force $\mathbf{F}%
_{L}=q\mathbf{E+}q\mathbf{U}\times \mathbf{B}$, $\mathbf{E=}0$ in $S,$ $%
\mathbf{B}$ is along the $+z$ axis, $q\mathbf{U}\times \mathbf{B}$ is the
magnetic part of the Lorentz force seen by the charges co-moving with the
disk along the segment $AC$. The integral along the segment $AC$ is taken at
the same moment $t$. In the $S^{\prime }$ frame the usual treatments suppose
that the Lorentz force becomes $\mathbf{F}_{L}^{\prime }=q\mathbf{E}%
_{st}^{\prime }\mathbf{+}q\mathbf{U}^{\prime }\times \mathbf{B}_{st}^{\prime
}$, where the components of the 3D $\mathbf{E}_{st}^{\prime }$\textbf{\ }and
$\mathbf{B}_{st}^{\prime }$ are determined by the ST (\ref{kr}). Thus it is
argued in the conventional formulation that in $S^{\prime }$ the charges
experiences the fields $\mathbf{E}_{st}^{\prime }=\gamma _{V}\mathbf{\beta }%
_{V}\times c\mathbf{B}$ and $\mathbf{B}_{st}^{\prime }=\gamma _{V}\mathbf{B,}
$ where $V=\omega R$, $\mathbf{\beta }_{V}=(V/c)\mathbf{i}$ and $\gamma
_{V}=(1-\beta _{V}^{2})^{-1/2}$. Then only the $y^{\prime }$ component of
the force $\mathbf{F}_{L}^{\prime }$ remains and it is
\begin{equation}
F_{L,y}^{\prime }=-qcB\beta _{U}/\gamma _{V}(1-\beta _{U}\beta _{V}).
\label{fe1}
\end{equation}
Notice that the same relation can be obtained from the definition of the
4-force (the components) $K^{\mu }=(\gamma _{U}\mathbf{F\cdot U},\gamma _{U}%
\mathbf{F})$ and its LT. This gives $\gamma _{U}^{\prime }F^{\prime
2}=\gamma _{U}F^{2}$ whence the same $F_{L,y}^{\prime }$ is obtained. (This
happens here accidentally since $F_{L,y}^{\prime }$ is calculated along the $%
y$ axis and $\mathbf{E=}0$ in $S$. Generally the expression $\mathbf{F}%
_{L}^{\prime }=q\mathbf{E}_{st}^{\prime }\mathbf{+}q\mathbf{U}^{\prime
}\times \mathbf{B}_{st}^{\prime }$ and the expression obtained from the LT
of the 4-force will not give the same result.) In $S^{\prime }$ the velocity
(in units of c) $\beta _{U}^{\prime }$ of some point on the segment $AC$ is $%
\beta _{U}^{\prime }=(\beta _{U}-\beta _{V})/(1-\beta _{U}\beta _{V})$ and
the corresponding $\gamma _{U}^{\prime }$ is $\gamma _{U}^{\prime }=\gamma
_{V}\gamma _{U}(1-\beta _{U}\beta _{V})$. The emf is again given by the
integral of $F_{L,y}^{\prime }/q$ over the common $y,y^{\prime }$ axis
(along the segment $AC,$ $\mathbf{dl}^{\prime }\mathbf{=dl}$) taken again at
the same moment of time, $t^{\prime }=0$ ($y$ axis is orthogonal to the
relative velocity $\mathbf{V}$)
\begin{equation}
emf^{\prime }=\int_{AC}(F_{L,y}^{\prime }/q)dy.  \label{emf2}
\end{equation}
It is clear from the expression for the emf in $S$, Eq. (\ref{emf1}), and
the corresponding one for the emf in $S^{\prime }$, Eq. (\ref{emf2}),
together with Eq. (\ref{fe1}) that these electromotive forces, in general,
are not equal. Really
\begin{equation}
emf^{\prime }=(c^{2}B/\omega \gamma _{V}\beta _{V}^{2})[\beta _{V}^{2}-\ln
(1+\beta _{V}^{2})],  \label{m1}
\end{equation}
thus $emf^{\prime }\neq emf.$ Only in the limit $\beta _{U},\beta _{V}\ll 1$
$emf^{\prime }\simeq emf$. \emph{This result explicitly shows that the
standard formulation is not relativistically correct formulation.}

Let us now consider the same example in the invariant relativistic
electrodynamics. In the tensor formalism the invariant Lorentz force $K^{a}$
is investigated in Ref. 11 Sec. 6.1. In terms of $F^{ab}$ it is $%
K^{a}=(q/c)F^{ab}u_{b},$ where $u^{b}$ is the 4-velocity of a charge $q$. In
the general case of an arbitrary spacetime and when $u^{a}$ is different
from $v^{a}$ (the 4-velocity of an observer who measures $E^{a}$ and $B^{a}$%
), i.e. when the charge and the observer have distinct world lines, $K^{a}$
can be written in terms of $E^{a}$ and $B^{a}$ as a sum of the $v^{a}$ -
orthogonal component, $K_{\perp }^{a}$, and $v^{a}$ - parallel component, $%
K_{\parallel }^{a}$, $K^{a}=K_{\perp }^{a}+K_{\parallel }^{a}.$ $K_{\perp
}^{a}$ is
\begin{equation}
K_{\perp }^{a}=(q/c^{2})\left[ \left( v^{b}u_{b}\right) E^{a}+c\widetilde{%
\varepsilon }^{a}\!_{bc}u^{b}B^{c}\right]  \label{kaokom}
\end{equation}
and $\widetilde{\varepsilon }_{abc}\equiv \varepsilon _{dabc}v^{d}$ is the
totally skew-symmetric Levi-Civita pseudotensor induced on the hypersurface
orthogonal to $v^{a}$, while
\begin{equation}
K_{\parallel }^{a}=(q/c^{2})\left[ \left( E^{b}u_{b}\right) v^{a}\right] .
\label{kapar}
\end{equation}
Speaking in terms of the prerelativistic notions one can say that $K_{\perp
}^{a}$, Eq. (\ref{kaokom}), plays the role of the usual Lorentz force lying
on the 3D hypersurface orthogonal to $v^{a}$, while $K_{\parallel }^{a}$,
Eq. (\ref{kapar}), is related to the work done by the field on the charge.
However \emph{in the invariant SR only both components together, that is, }$%
K^{a}$, \emph{does have definite physical meaning and }$K^{a}$\emph{\
defines the Lorentz force both in the theory and in experiments.} Of course $%
K^{a}$, $K_{\perp }^{a}$ and $K_{\parallel }^{a}$ are all 4D quantities
defined without reference frames, the AQs, and the decomposition of $K^{a}$
is an observer independent decomposition. Then we define the emf also as an
invariant 4D quantity, the Lorentz scalar,
\begin{equation}
emf=\int_{\Gamma }(K^{a}/q)dl_{a},  \label{emfi}
\end{equation}
where $dl_{a}$ is the infinitesimal spacetime length and $\Gamma $ is the
spacetime curve. Let the observers are at rest in the $S$ frame, $v^{\mu
}=(c,0,0,0)$ whence $E^{0}=B^{0}=0$; the $S$ frame is the rest frame of
``fiducial'' observers, the $\gamma _{0}$ - frame with the $\left\{ \gamma
_{\mu }\right\} $ basis. Thus the components of the 4-vectors in the $%
\left\{ \gamma _{\mu }\right\} $ basis are $E^{\mu }=(0,0,0,0)$, $B^{\mu
}=(0,0,0,B)$, $u^{\mu }=(\gamma _{U}c,\gamma _{U}U=\gamma _{U}\omega y,0,0)$%
, $dl^{\mu }=(0,0,dl^{2}=dy,0)$. Thence $K_{\parallel }^{a}=0$, $K_{\perp
}^{0}=K_{\perp }^{1}=K_{\perp }^{3}=0$, $K_{\perp }^{2}=\gamma _{U}qUB$.
When all quantities in Eq. (\ref{emfi}) are written as CBGQs in the $S$
frame with the $\left\{ \gamma _{\mu }\right\} $ basis we find
\begin{equation}
emf=(c^{2}B/\omega )\{1-[1-(\omega R/c)^{2}]^{1/2}\},  \label{f3}
\end{equation}
which for $\omega R/c\ll 1$ becomes the usual expression $emf=\omega
R^{2}B/2 $ as in Eq. (\ref{emf1}). Since the expression (\ref{emfi}) is
independent of the chosen reference frame and of the chosen system of
coordinates in it we shall get the same result, Eq. (\ref{f3}), in the
relatively moving $S^{\prime }$ frame as well;
\begin{eqnarray}
emf &=&\int_{\Gamma (in\ S)}(K^{\mu }/q)dl_{\mu }=\int_{\Gamma (in\
S^{\prime })}(K^{\prime \mu }/q)dl_{\mu }^{\prime }  \notag \\
&=&(c^{2}B/\omega )\{1-[1-(\omega R/c)^{2}]^{1/2}\}.  \label{ef2}
\end{eqnarray}
This can be checked directly performing the LT of all 4-vectors as CBGQs
from $S$ to $S^{\prime }$ including the transformation of $v^{\mu }\gamma
_{\mu }$. Obviously \emph{the approach with Lorentz invariant 4D quantities
gives the relativistically correct answer }in an enough simple and
transparent way. From the viewpoint of the geometric approach the agreement
with the usual approach exists only in the frame of the ``fiducial''
observers and when $V\ll c$.\bigskip \medskip

\noindent \textbf{5.\ SUMMARY AND CONCLUSIONS}\bigskip

The covariance of the ME is cosidered to be a cornerstone of the modern
relativistic field theories, both classical and quantum. Einstein$^{(4)}$
derived the ST of the 3D $\mathbf{E}$ and $\mathbf{B}$ assuming that the ME
with $\mathbf{E}$ and $\mathbf{B}$ must have the same form in all relatively
moving inertial frames of reference. In Einstein's formulation of SR$^{(4)}$
the principle of relativity is a fundamental postulate that is supposed to
hold for all physical laws including those expressed by 3D quantities, e.g.,
the ME with the 3D $\mathbf{E}$ and $\mathbf{B.}$ This derivation is
discussed in detail in Ref. 11. The results presented in this paper
substantially change generally accepted opinion about the covariance of the
ME exactly proving in the geometric algebra and tensor formalisms that \emph{%
the usual ME} ((\ref{MEC}), or (\ref{H2}), or (\ref{me})) \emph{change their
form upon the LT} (see Eq. (\ref{mcr}) with Eq. (\ref{anu}), or Eq. (\ref
{ehbc}) with Eq. (\ref{act}), or Eq. (\ref{abc}) with $a^{\prime 0}$ from
Eq. (\ref{anu})). \emph{It is also proved that the ST of the} \emph{ME} (see
Eqs. (\ref{rtr}) and (\ref{EBC}), or Eq. (\ref{gc}), or Eq. (\ref{cm})),
\emph{which leave unchanged the form of the ME, actually have nothing in
common with the LT of the usual ME.} The difference between the LT of the
ME, e.g., Eq. (\ref{mcr}) with Eq. (\ref{anu}), and their ST, e.g., Eqs. (%
\ref{rtr}) and (\ref{EBC}), is essentially the same as it is the difference
between the LT of the electric and magnetic fields (see Eqs. (\ref{nle}) and
(\ref{nlb}), or Eq. (\ref{eh}), or Eq. (\ref{ebcr})) and their ST (see Eqs. (%
\ref{ce}) and (\ref{B}), or Eq. (\ref{es}), or Eq. (\ref{kr})). This last
difference is proved in detail in Refs. 1, 2 and that proof is only briefly
repeated in this paper. All this together reveals that, contrary to the
generally accepted opinion, \emph{the principle of relativity does not hold
for physical laws expressed by 3D quantities (a fundamental achievement).}
\emph{A 3D quantity cannot correctly transform upon the LT and thus it does
not have an independent physical reality in the 4D spacetime; it is not the
same quantity for relatively moving observers in the 4D spacetime} (see
also, e.g., Figs. 3. and 4. in Ref. 11, and Ref. 12). Since the usual ME
change their form upon the LT they cannot describe in a relativistically
correct manner the experiments that test SR, i.e., the experiments in which
relatively moving observers measure \emph{the same 4D physical quantity}.
Therefore the new field equations with geometric 4D quantities are
constructed in geometric algebra formalism with 1-vectors $E$ and $B$ (Eqs. (%
\ref{deb}) and (\ref{maeb})), and with bivectors $E_{Hv}$ and $B_{Hv}$ (Eqs.
(\ref{Nf}) and (\ref{nh}) with (\ref{D})), and also in the tensor formalism
with 4-vectors $E^{a}$ and $B^{a}$ (Eq. (\ref{em2})); the Lorentz invariant
field equations in the tensor formalism are already presented in Refs. 11,
20. All quantities in these geometric equations are independent of the
chosen reference frame and of the chosen system of coordinates in it. When
the $\gamma _{0}$ - frame with the $\left\{ \gamma _{\mu }\right\} $ basis
is chosen, in which the observers who measure the electric and magnetic
fields are at rest, then all mentioned geometric equations become the usual
ME. This result explicitly shows that the correspondence principle is
naturally satisfied in the invariant SR. However, as seen here, \emph{the
description with 4D geometric quantities} \emph{is correct not only in the} $%
\gamma _{0}$ - \emph{frame with the }$\left\{ \gamma _{\mu }\right\} $ \emph{%
basis} \emph{but in all other relatively moving frames and it holds for any
permissible choice of coordinates. We conclude from the results of this
paper that geometric 4D quantities, defined without reference frames, i.e.,
the AQs, or as CBGQs, have an independent physical reality and the
relativistically correct physical laws are expressed in terms of such
quantities.} The principle of relativity is automatically satisfied with
such quantities whereas in the standard formulation of SR it is postulated
outside the mathematical formulation of the theory. We see that the role of
the principle of relativity is substantially different in the Einstein
formulation of SR and in the invariant SR. The results of this paper clearly
support the latter one. Furthermore we note that all observer independent
quantities, i.e., the AQs, introduced here and the field equations written
in terms of them hold in the same form both in the flat and curved
spacetimes. The results obtained in this paper will have important and
numerous consequences in all relativistic field theories, classical and
quantum. Some of them will be soon examined. \medskip \bigskip

\noindent \textbf{ACKNOWLEDGEMENTS\bigskip }

\noindent I am grateful to Professor Larry Horwitz for his continuos
interest, support and useful comments and to Professor Alex Gersten and
other participants of the IARD 2004 Conference for interesting discussions.
\medskip \bigskip

\textbf{REFERENCES}\bigskip

\noindent 1. T. Ivezi\'{c}, \textit{Found. Phys.} \textbf{33}, 1339 (2003);
hep-th/0302188.

\noindent 2. T. Ivezi\'{c}, physics/\textit{0304085}.

\noindent 3. H.A. Lorentz, \textit{Proceedings of the Academy of Sciences of
Amsterdam},

6 (1904), in W. Perrett and G.B. Jeffery, in \textit{The Principle of
Relativity}

(Dover, New York).

\noindent 4. A. Einstein, \textit{Ann. Physik.} \textbf{17}, 891 (1905), tr.
by W. Perrett and G.B.

Jeffery, in \textit{The Principle of Relativity} (Dover, New York).

\noindent 5. J.D. Jackson, \textit{Classical Electrodynamics} (Wiley, New
York, 1977)

2nd edn.; L.D. Landau and E.M. Lifshitz, \textit{The Classical Theory of }

\textit{Fields, }(Pergamon, Oxford, 1979) 4th edn.;

\noindent 6. C.W. Misner, K.S.Thorne, and J.A. Wheeler, \textit{Gravitation}
(Freeman, San

Francisco, 1970).

\noindent 7. D. Hestenes, \textit{Space-Time Algebra }(Gordon and Breach,
New York, 1966);

\textit{Space-Time Calculus; }available at: http://modelingnts.la.
asu.edu/evolution.

html; \textit{New Foundations for Classical Mechanics }(Kluwer Academic

Publishers, Dordrecht, 1999) 2nd. edn.; \textit{Am. J Phys.} \textbf{71},
691 (2003).

\noindent 8. C. Doran, and A. Lasenby, \textit{Geometric algebra for
physicists }

(Cambridge University Press, Cambridge, 2003).

\noindent 9. B. Jancewicz, \textit{Multivectors and Clifford Algebra in
Electrodynamics}

(World Scientific, Singapore, 1989).

\noindent 10. T. Ivezi\'{c}, physics/\textit{0305092}.

\noindent 11. T. Ivezi\'{c}, \textit{Found. Phys.} \textbf{31}, 1139 (2001).

\noindent 12. T. Ivezi\'{c}, \textit{Found. Phys. Lett.} \textbf{15}, 27
(2002); physics/0103026; physics/

0101091.

\noindent 13. T. Ivezi\'{c}, physics/\textit{0311043}.

\noindent 14. M. Riesz, \textit{Clifford Numbers and Spinors}, Lecture
Series No. 38,

The Institute for Fluid Dynamics and Applied Mathematics,

University of Maryland (1958).

\noindent 15. T. Ivezi\'{c}, hep-th/0207250; hep-ph/0205277.

\noindent 16. A. Einstein, \textit{Ann. Physik }\textbf{49,} 769 (1916), tr.
by W. Perrett and G.B.

Jeffery, in \textit{The Principle of Relativity }(Dover, New York).

\noindent 17. H.N. N\'{u}\~{n}ez Y\'{e}pez, A.L. Salas Brito, and C.A.
Vargas, \textit{Revista}

\textit{Mexicana de F\'{i}sica} \textbf{34}, 636 (1988).

\noindent 18. T. Matolcsi, \textit{Spacetime without Reference Frames}
(Akad\'{e}miai

Kiad\'{o}, Budapest, 1993).

\noindent 19. D. Hestenes and G. Sobczyk, \textit{Clifford Algebra to
Geometric Calculus}

(Reidel, Dordrecht, 1984).

\noindent 20. T. Ivezi\'{c}, \textit{Annales de la Fondation Louis de Broglie%
} \textbf{27}, 287 (2002).

\noindent 21. S. Esposito, \textit{Found. Phys.} \textbf{28}, 231 (1998).

\noindent 22. T. Ivezi\'{c}, \textit{Found. Phys. Lett.} \textbf{12}, 105
(1999); \textit{Found. Phys. Lett.} \textbf{12},

507 (1999).

\noindent 23. R.M. Wald, \textit{General Relativity} (The University of
Chicago Press,

Chicago, 1984).

\noindent 24. M. Ludvigsen, \textit{General Relativity,} \textit{A Geometric
Approach }

(Cambridge University Press, Cambridge, 1999); S. Sonego and

M.A. Abramowicz, \textit{J. Math. Phys.} \textbf{39}, 3158 (1998); D.A. T.
Vanzella,

G.E.A. Matsas, H.W. Crater, \textit{Am. J. Phys.} \textbf{64}, 1075 (1996).

\noindent 25. F. Rohrlich, \textit{Nuovo Cimento B} \textbf{45}, 76 (1966).

\noindent 26. A. Gamba, \textit{Am. J. Phys.} \textbf{35}, 83 (1967).

\noindent 27. G. Saathoff et al., \textit{Phys. Rev. Lett.} \textbf{91,}
190403 (2003); H. M\"{u}ller et al.,

\textit{Phys. Rev. Lett.} \textbf{91,} 020401 (2003); P. Wolf et al.,
\textit{Phys. Rev. Lett.}

\textbf{90,} 060402 (2003).

\noindent 28. W.K.H. Panofsky and M. Phillips, \textit{Classical electricity
and magnetism,}

2nd edn. (Addison-Wesley, Reading, Mass., 1962).

\noindent 29. L. Nieves, M. Rodriguez, G. Spavieri and E. Tonni, \textit{%
Nuovo Cimento B}

\textbf{116}, 585 (2001).

\end{document}